\documentstyle[12pt]{article}
\begin{document}
\def\ch#1{#1}
\def\ket#1{| #1 \rangle}
\def\bracket#1#2{\langle #1 | #2\rangle}
\baselineskip 0.2cm
\normalsize
\bigskip
\begin{center}
{\huge Polarons}
\end{center}
\vskip 0.5truecm
\begin{center}
{\large J.T.\ Devreese
\bigskip

{\it Universiteit Antwerpen, Belgium,
\medskip

and Technische Universiteit Eindhoven, The Netherlands}}

\bigskip
Encyclopedia of Applied Physics, Vol. 14, pp. 383 -- 409 (1996)
\copyright 1996 Wiley-VCH Publishers, Inc.
\end{center}

\vskip 0.5truecm
\begin{center}
{Table of Contents}
\end{center}

{\small

{\bf \hspace {0.3cm} Introduction}
\medskip

{\bf         1 The Polaron Concept and the Fr\"ohlich Polaron}
\medskip

{1.1 The concept}
\medskip

{1.2 Standard Fr\"ohlich-polaron theory.
Self energy and effective mass of the polaron}
\medskip






{1.3 Fr\"ohlich polarons in 2D}
\medskip



{1.4 Fr\"ohlich polarons in a magnetic field}
\medskip







{1.5 The bound polaron}
\medskip

{\bf 2       The Small Polaron}
\medskip

{2.1 The small-polaron concept. Role of localization}
\medskip

{2.2 Standard small-polaron theory}
\medskip


{2.3 Experimental evidence}
\medskip

{\bf 3          Bipolarons and Polaronic Excitons}
\medskip

{3.1 Fr\"ohlich bipolarons}
\medskip

{3.2 Polaronic excitons}
\medskip

{3.3 Localized bipolarons}
\medskip

{\bf 4        Spin Polarons}
\medskip

{\bf 5        Bipolarons and  High-Temperature  Superconductivity}
\medskip

{\bf 6      Further Developments of the Polaron Concept}
\medskip

{6.1        Polarons and bipolarons in polymers. Solitons}
\medskip

{6.2        Modelling systems using the polaron concept}
\medskip


{\bf \hspace {0.3cm} List of Works Cited}
\medskip

{\bf \hspace {0.3cm} Further Reading}
}

\newpage
\begin{center}
{Abstract}
\end{center}
A conduction electron (or hole) together with its self-induced
polarization in a polar semiconductor or an ionic crystal forms a
quasiparticle, which is called a polaron.
The polaron concept is of interest, not only because it describes the
particular physical properties of charge carriers in polarizable solids
but also because it constitutes an interesting field
theoretical model consisting of a fermion interacting with a scalar
boson field.
The early work on polarons was concerned with general theoretical
formulations and approximations, which now constitute the standard
polaron theory and with experiments on cyclotron resonance and transport
properties.
Because of the more recent interest in the two-dimensional electron gas,
the study of the polaron in two dimensions became important.
Again cyclotron resonance, and therefore the behaviour of polarons in
magnetic fields, was a key issue.
When two electrons (or two holes) interact with each other
simultaneously through the Coulomb force and via the
electron-phonon-electron interaction, either two independent polarons can
occur or a bound state of two polarons --- a bipolaron --- can arise.
Bipolarons have been
considered to possibly  play a role in high-T$_C$ superconductivity.
Originally the polaron was mainly studied to
describe the polar interaction between an electron and the longitudinal
optical phonons. The polaron concept has been extended to several
systems where one or many fermions interact with a bath of bosons, e. g.,
\begin{itemize}
\item small polaron,
\item piezopolaron,
\item polaronic exciton,
\item spin --- or magnetic --- polaron,
\item ``ripplonic polaron'',
\item ``plasmaron'',
\item ``hydrated polarons'',
\item electronic polaron
\end{itemize}
etc.
Polarons in conducting polymers and in fullerenes also constitute
an emerging subject.

\newpage

\section*{Introduction}
The polaron concept is of interest, not only because it describes the
particular physical properties of an electron in polar crystals and
ionic semiconductors but also because it is an interesting field
theoretical model consisting of a fermion interacting with a scalar
boson field.

The early work on polarons was concerned with general theoretical
formulations and approximations, which now constitute the standard
theory and with experiments on cyclotron resonance and transport
properties.

Because of the more recent interest in the two-dimensional electron gas
(2DEG) the study of the polaron in two dimensions became important.
A cyclotron resonance, and therefore the behaviour of polarons in
magnetic fields, was a key issue.

\section{\ch{The Polaron Concept and the Fr\"ohlich \\ Polaron}}
\subsection{The concept}
A conduction electron (or hole) together with its self-induced
polarization in a polar semiconductor or an ionic crystal forms a
quasiparticle, which is called a {\em POLARON} (Fig.\,1).
The physical
properties of the polaron differ from those of the band-electron. In
particular the polaron is characterized by its binding or (self-)
energy, effective mass and by its response to
external electric and magnetic fields (e. g. mobility and impedance).
The general polaron concept was introduced by Landau (1933) in a paper of
about one page. Subsequently  Landau and Pekar [see (Pekar, 1951)]
investigated the self-energy and the effective mass
of the polaron, for what was shown by Fr\"{o}hlich (1954) to
correspond to the adiabatic or strong-coupling regime.

The early work on polarons was devoted to the interaction between a
charge carrier (electron, hole) and the long-wavelength optical phonons.
The (now standard) field-theoretical Hamiltonian describing this
interaction was derived by Fr\"{o}hlich:
$$
H=\frac{{\bf p}^{2}}{2m_b}   + \sum_{{\bf k}}    \hbar \omega_{LO} a^{+}_{{\bf k}}  a_{{\bf k}}
+\sum_{{\bf k}} (V_{k}   a_{{\bf k}}  e^{i {\bf {\bf k} \cdot r}} + h.c.),
\hspace*{0.7truecm}
(1a)
$$
where ${\bf  r}$ is the position coordinate operator of the electron
with band mass $m_b$, ${\bf p}$ is its canonically conjugate momentum
operator; $a^{\dagger}_{{\bf k}}$ and $a_{{\bf k}}$ are the creation (and annihilation)
operators for longitudinal optical phonons of wave vector ${\bf k}$ and
energy $\hbar \omega_{LO}$. The $V_{k}$ are Fourier components of the
electron-phonon interaction
$$
V_{k}=-i \frac{\hbar \omega_{LO}}{k}
\left( \frac{4\pi \alpha}{V}        \right)^{\frac{1}{2}}
\left( \frac{\hbar}{2m_b\omega_{LO}}   \right)^{\frac{1}{4}},
\hspace*{0.7truecm}
(1b)
$$
$$
\alpha= \frac{e^{2}}{\hbar c}  \sqrt{\frac{m_bc^{2}}{2\hbar \omega_{LO}}}
\left(\frac{1}{\varepsilon_{\infty}}-\frac{1}{\varepsilon_{0}}\right)
\hspace*{0.7truecm}
(1c)$$
$\alpha$ is called the Fr\"ohlich coupling constant,
$\varepsilon_{\infty}$ and $\varepsilon_{0}$ are respectively the
electronic and the static dielectric constant of the polar crystal.
In   \ch{Table 1} a list is given of the coupling constants for a number of
crystals.

In deriving the form of $V_{k}$, Eqs.\,(1b) and (1c), it was
assumed that (i) the spatial extension of the polaron is large
compared to the lattice parameters of the solid
(``continuum'' approximation), (ii)
spin and relativistic effects can be neglected, (iii) the
band-electron has parabolic dispersion, (iv) in conjunction with the
first approximation it is also assumed that the LO-phonons of
interest for the interaction, are the long-wavelength phonons with
constant frequency $\omega_{LO}$.

The original concept of the polaron, as discussed above, has been
generalized over the years to include polarization fields other
than the LO-phonon field: the acoustical phonon field, the exciton
field, etc. \ldots. For some materials the continuum approximation is
not appropriate as far as the polarization is confined to a region of the
order of a unit cell; the so-called ``small polaron'' is a more
adequate quasiparticle in that case (see section \ref{secsp}).

It is customary to use the term ``Fr\"{o}hlich-polaron'' or \ch{``large
polaron''} for the
quasiparticle consisting of the electron (or hole) and the polarization
due to the LO-phonons. The term Landau- or Pekar- or
Landau-Fr\"{o}hlich-polaron would be more appropriate.

\ch{In this first section the properties of Fr\"ohlich polarons are
reviewed; other realisations of the polaron concept (small polarons,
spin polarons etc\ldots) are discussed in later sections. For the reviews on
polarons see Kuper and Whitfield (1963),
Appel (1968), Devreese (1972), Devreese and Peeters (1984).}
\subsection{\ch{Standard Fr\"ohlich-polaron theory.
Self energy and effective mass of the polaron}}

Historically the first studies on polarons [the ``Russian work'':
Landau (1933), Pekar (1951)]
were based on a ``Produkt-Ansatz'' for the polaron wave-function
$$
\ket{\Phi} = \ket{\Psi({\bf r})}   \ket{\mbox {\rm field}}
= \ket{\Psi({\bf r})}\ket{f}
\hspace*{0.6truecm}
(2a) $$
where $\ket{\Psi(\bf r)}$ is the electron-wave function. The field wave
function parametrically depends on the electron wave function.
Fr\"{o}hlich showed that the approximation (2a) leads to results which
are only valid as $\alpha \rightarrow \infty$ (the strong-coupling regime).
It should be pointed out that a more systematic analysis of the
strong-coupling polarons based on canonical transformations of
the Hamiltonian (1a) was performed in pioneering investigations by Bogolubov
(1950), Bogolubov and Tyablikov (1949), Tyablikov (1951).

Fr\"ohlich also found that Eq.\,(2a) is a poor Ansatz to represent actual
crystals \ch{which have} $\alpha$-values typically ranging from
$\alpha=0.02$ (InSb) to $\alpha \sim 3$ to $4$ (alkali halides).
This work showed the need for a weak-coupling theory  of
the polaron which in fact was provided originally by Fr\"{o}hlich (1954).
However it turns out that for $\alpha \approx 3$
perturbation expansions in \ch{powers of} $\alpha$ are not always sufficient; therefore
an intermediate-coupling or --- better --- an all-coupling theory was
necessary.

In what follows the key concepts of the standard \ch{Fr\"ohlich}-polaron
theories \ch{are reviewed}.
\subsubsection{Strong coupling}
The
Produkt-Ansatz (2a) --- or Born-Oppenheimer approximation --- implies that
the electron adiabatically follows the motion of the atoms (Pekar, 1951).
With the use of the canonical transformation
$$
S=\exp \left[-\sum_{{\bf k}}   \left(\frac{V_{k}^{*} \rho_{{\bf k}}^{*}}
{e\hbar \omega_{LO}}a_{{\bf k}} -h.c.\right)\right],
\hspace*{0.5truecm}
(2b)
$$
where
$$
\rho_{{\bf k}} = e <\Psi({\bf r})|e^{i {\bf k \cdot r}}|\Psi({\bf r})>,
\hspace*{0.7truecm}
(2c)$$
it leads to $\ket{f}=S\ket{0}$. The ket $\ket{0}$ describes the vacuum
state. With Eqs.\,(2) and a Gaussian trial function for $\ket{\Psi({\bf r})}$,
the groundstate energy of the polaron $E_{0}$ (calculated with the
energy of the uncoupled electron-phonon system as zero energy) takes the
form:
$$
E_{0}= -\frac{\alpha^{2}}{3 \pi} \hbar \omega_{LO} = -0.106 \alpha^{2} \hbar
\omega_{LO}.
\hspace*{0.3truecm}
(3a)$$
At strong coupling, the polaron is characterized by Franck-Condon (F.~C.)
excited states, which correspond to  excitations of the electron in
the potential adapted to the groundstate. The energy of the lowest F.~C.
state is, within the Produkt-Ansatz:
$$
E_{FC} = \frac{\alpha^{2}}{9 \pi} \hbar \omega_{LO}= 0.0354 \alpha^{2} \hbar
\omega_{LO}.
\hspace*{0.4truecm}
(3b)
$$
If the lattice polarization is allowed to relax or adapt to the
electronic distribution of the excited electron (which itself then
adapts its wave function to the new potential, etc. \ldots leading to a
self-consistent final state), the so-called relaxed excited state (R.~E.~S.)
results (Pekar, 1951). Its energy is (Evrard, 1965; Devreese, Evrard, 1966):
$$
E_{RES} = -0.041 \alpha^{2} \hbar \omega_{LO}.
                \hspace*{0.3truecm}
                (3c)$$
In fact, both the F.~C. state and the R.~E.~S. lie in the
continuum
\footnote{The R.~E.~S. lies in the continuum of so-called scattering
states, not indicated in the figure.}
and, strictly speaking, are resonances (Fig.\,2).

The strong-coupling mass of the polaron, resulting again from the
approximation (2), is given as:
$$
\frac{m^{*}}{m_b}= 1+ 0.0200 \alpha^{4}.
\hspace*{0.3truecm}
(3d)
$$
More rigorous strong-coupling expansions for $E_{0}$ and $m^{*}$ have
been presented in the literature (Miyake, 1975):
$$
\frac{E_{0}}{\hbar \omega_{LO}} = - 0.108513 \alpha^{2}  -2.836                                    ,
\hspace*{0.3truecm}
(4a)$$
$$
\frac{m^{*}}{m_b} = 1+ 0.0227019 \alpha^{4}.
\hspace*{0.3truecm}
(4b)$$
The main significance of the strong-coupling theory is that it allows to
test ``all-coupling'' theories in the limit $\alpha \rightarrow \infty$.
\ch{From the formal point of view a deeper study of the status and
uniqueness of the strong-coupling solutions was undertaken by Lieb
(1977).}
\subsubsection{Weak coupling}
Fr\"{o}hlich (1954)
has provided the first weak-coupling perturbation-theory results:
$$
E_{0}= -\alpha \hbar \omega_{LO}
\hspace*{0.3truecm}
(5a)$$
and
$$
m^{*}= \frac{m_b}{1-\alpha/6}.
\hspace*{0.3truecm}
(5b)$$
Inspired by the work of Tomonaga,  Lee {\it et al.} (1953) (usually cited
as L.~L.~P.)
have derived (5a) and $m^{*}=m_b (1+\alpha/6)$ from an elegant canonical
transformation formulation. In fact, they perform two successive
canonical transformations:
$$
S_{1}=\exp\left[\frac{i}{\hbar}({\bf P} -\sum_{{\bf k}} \hbar {\bf k} a_{{\bf k}}^{\dagger}
a_{{\bf k}})   {\bf \cdot r}\right],
\hspace*{0.3truecm}
(6a)$$
where ${\bf P}$ is the polaron total-momentum operator. The canonical
transformation (6a) formally eliminates the electron operators from the
Hamiltonian.

The second canonical transformation is of the ``displaced-oscillator''
form:
$$
S_{2}= \exp\left[\sum_{{\bf k}}   (a_{{\bf k}}^{\dagger} f_{{\bf k}}  -a_{{\bf k}}f_{{\bf k}}^{*})\right].
\hspace*{0.3truecm}
(6b)$$
The $f_{{\bf k}}$ are treated as variational functions. The physical
significance of Eq.\,(6b) is that it ``dresses'' the electron with the virtual
phonon field which  describes the polarization.

The L.~L.~P. approximation has often been called ``intermediate-coupling
theory''. However its range of validity is in principle not larger than
that of the weak-coupling approximation. The  main significance of the
L.~L.~P. approximation, is in the elegance of the
canonical transformations $S_{1}$ and $S_{2}$, together with the fact
that it puts the Fr\"{o}hlich result on a variational basis.

In his basic article on polarons, Feynman (1955) found the following
higher-order weak-coupling expansions:
$$
\frac{E_{0}}{\hbar \omega_{LO}} = -\alpha - 0.0123 \alpha^{2} - 0.00064
\alpha^{3} - \ldots
\>(\alpha \rightarrow 0),
\hspace*{0.3truecm}
(7a)
$$
$$
\frac{m^{*}}{m_b}= 1+ \frac{\alpha}{6} + 0.025 \alpha^{2} + \ldots
\>(\alpha \rightarrow 0).
\hspace*{0.3truecm}
(7b)$$
Since then a lot of theoretical work was devoted to obtain more exact
coefficients in this expansion.
H\"ohler and M\"ullensiefen (1959) calculated the coefficients for
$\alpha^2$ to be $-0.016$ in the energy and $0.0236$ in the
polaron mass. R\"oseler (1968) found the analytical
expressions for the above-mentioned coefficients:
$2\ln (\sqrt{2} +1) - {3 \over 2}\ln 2 -{\sqrt{2} \over 2} \approx
-0.01591962$ and ${4 \over 3} \ln (\sqrt{2} +1) -{2 \over 3} \ln 2
-{5 \sqrt{2} \over 8} + {7 \over 36} \approx 0.02362763$, respectively.
At present the following most accurate higher-order weak-coupling expansions
are known: for the energy (Smondyrev, 1986; Selyugin and Smondyrev, 1989)
$$
\frac{E_{0}}{\hbar \omega_{LO}} = -\alpha - 0.0159196220 \alpha^{2}
- 0.000806070048 \alpha^{3} - \ldots,
\hspace*{0.3truecm}
(7c)
$$
and for the polaron mass (R\"oseler, 1968)
$$
\frac{m^{*}}{m_b}= 1+ \frac{\alpha}{6} + 0.02362763\alpha^{2} + \ldots,
\hspace*{0.3truecm}
(7d)$$
which are a striking illustration of the heuristical value of Feynman's
path-integral approach in the polaron theory.

\subsubsection{All-coupling theory. Feynman path integral.}
In the early fifties H. Fr\"{o}hlich gave a seminar at Caltech. In this
seminar he discussed the weak-coupling polaron mass as derived by him:
$m^{*}=m_b/(1-\alpha/6)$. He suggested that, if the electron-phonon
coupling could be accurately treated for intermediate coupling (in
particular around $\alpha \approx 6$) this might lead to new insights in
the theory of superconductivity (this was before BCS) (Feynman, 1973).
Feynman was among the audience. He went to the library to study one of
Fr\"{o}hlich's papers on polarons, (Fr\"ohlich, 1954). There he got the
idea to formulate the polaron problem into the Lagrangian form, of
quantum mechanics and then eliminate the field oscillators, "\ldots in
exact analogy to Q.~E.~D. \ldots (resulting in) \ldots a sum over all
trajectories \ldots". The resulting path integral is of the form
(Feynman, 1955):
$$
\bracket{0,\beta}{0,0}\!=\!\int\!{\cal D}{\bf r}(\tau) \exp \left[-\frac{1}{2}
\int_{0}^{\beta} {\bf {\dot r}} ^{2} d \tau\!+\!
\frac{\alpha}{2^{\frac{3}{2}}}     \int_{0}^{\beta}\!\int_{0}^{\beta}
\frac{e^{-|\tau - \sigma|}}{|{\bf r}(\tau)- {\bf r}(\sigma)|}
d\tau d\sigma\right],
\hspace*{0.3truecm}
(8a)$$
where $\beta = 1/(k_B T)$.
This path integral (8a) has a great intuitive appeal: it shows the polaron
problem as an equivalent one-particle problem in which the interaction,
non-local in time \ch{or ``retarded''}, is between the electron and itself.
Subsequently
Feynman showed (in fact to \ch{M.~Baranger}) how the variational principle
of quantum mechanics could be adapted to the path integral and
introduced  a quadratic trial action (again non-local in time) to
simulate Eq.\,(8a).

\ch{It may be noted,
that the elimination of the phonon
field (or the boson field in general) introduced through Eq.\,(8a)
by Feynman
has found many applications, e.~g. in the study of dissipation phenomena}

Applying the variational principle \ch{for path integrals} resulted in an upper bound for the
polaron self-energy at all $\alpha$, which at weak and strong coupling
gave accurate limits.
Feynman obtained the smooth interpolation between weak and strong coupling
(for the groundstate energy).
It is worthwhile to
give the asymptotic expansions of Feynman's polaron theory. In the
weak-coupling limit, they are given above by Eqs. (7a) and (7b). In another ---
strong-coupling limit --- Feynman found for the energy:

$$
\frac{E_0}{\hbar \omega_{LO}}
\equiv\frac{E_{3D}(\alpha)}{\hbar \omega_{LO}}
=-0.106 \alpha^{2} -2.83 - \ldots
\hspace*{0.3truecm}
(\alpha
\rightarrow \infty).
\hspace*{0.3truecm}
(9a)$$
and for the polaron mass:
$$
\frac{m^{*}}{m_b}
\equiv\frac{m^*_{3D}(\alpha)}{m_b}
=  0.0202 \alpha^{4} + \ldots
\hspace*{0.3truecm}
(\alpha \rightarrow \infty).
\hspace*{0.3truecm}
(9b)
$$
Over the years the Feynman model for the polaron has remained \ch{in many
respects} the most
successful approach to this problem. It is also remarkable that, despite
several efforts, no equivalent Hamiltonian formulation of this path
integral approach has been realised. [In one case (Yamazaki, 1983) the formal
structure of the theory was reobtained in a Hamiltonian formulation---be
it very artificial---but no variational principle \ch{leading to an
upper bound for the energy} could be found.]
The theory of polarons in semiconductors with degenerate bands and
of statistical ensembles of polarons, related to many-polaron problems,
is reviewed from the unified point of view of the path
integration over the Wick symbols by Fomin and Pokatilov (1988).

\subsubsection{Response properties of Fr\"ohlich polarons. Mobility and \\
optical properties}
The transport properties of polar and ionic solids are influenced by the
polaron coupling. Intuitively one expects that the mobility of large
polarons will be inversely proportional to the number of real phonons
present in the crystal:
$$
\mu \sim \hbox{\rm e}^{\hbar\omega_{LO}/kT}.
\hspace*{0.3truecm}
(10a)
$$
The first to point out the typical behaviour (characteristic for weak
coupling) of Eq.\,(10a) was Fr\"ohlich (1937).
Kadanoff (1963) provided a derivation of the weak-coupling,
low-temperature mobility starting from the Boltzmann equation; his
result is
$$
\mu =
\frac{4}{3\pi^{1/2}}\frac{e}{m^*\alpha\omega_{LO}}G(z)
\hbox{\rm e}^{\hbar\omega_{LO}/kT}
\left(\frac{kT}{\hbar\omega_{LO}}\right)^{1/2},
\hspace*{0.3truecm}
(10b)
$$
where $G(z)$, defined by Howarth and Sondheimer (1953),
is of order 1.

Feynman {\it et al.} (1962) (usually referred to as F.~H.~I.~P.) have
elaborated a framework allowing the
analysis of the response properties of a system, using path integrals.
For low temperatures they obtain the following expression for the dc
mobility:
$$
\mu
\equiv \mu_{3D}(\alpha)
=\frac{e}{2m^*\omega_{LO}\alpha}\frac{3}{2}\left(\frac{w}{v}\right)^3
\exp\left({\frac{v^2-w^2}{w^2v}}\right)
\hbox{\rm e}^{\hbar\omega_{LO}/kT} \frac{kT}{\hbar\omega_{LO}},
\hspace*{0.3truecm}
(10c)
$$
where $w,v$ are functions of $\alpha$ (also adequate for large $\alpha$)
deriving from the Feynman polaron model\footnote{It is noted that a misprint
occurred in this formula in the
{\it Encyclopedia of Physics} (Lerner, Trigg, 1991).}.
For the impedance of a polaron
$Z(\nu) \equiv Z_{3D}(\alpha, \nu)$ describing its response to the ac
electric field of frequency $\nu$, see Feynman {\it et al.} (1962).

Equation\,(10c) and related approximations allowed to explain the
experimental results of Brown and his co-workers [see (Brown, 1963)]
on alkali halides and silver halides, see e.~g. Fig.\,3, in the temperature
region where the electron-LO-phonon scattering is the dominant process
(at low temperatures $T < 50$K the impurity scattering starts to prevail).
Equation\,(10c)
presupposed a ``drifted-Maxwellian'' velocity distribution of the
polarons; this limits the validity of the treatment theoretically.
However, because other scattering mechanisms than LO-phonon scattering
might ``conspire'' to induce a Maxwellian distribution, the range of
validity of Eq.\,(10c) seems to be larger than one would expect from
electron-LO-phonon interaction only. A review on the mobility of
Fr\"ohlich polarons, written from a unified point of view, was presented
by Peeters and Devreese (1984).

It was shown by the author and his co-authors [see (Devreese, 1972)]
how the optical absorption of Fr\"ohlich polarons, for all coupling,
can be calculated
starting from the F.~H.~I.~P.-scheme (which gives a derivation for the
impedance function). This work led to the following
expression for the optical absorption:
$$
\Gamma(\nu) \sim \frac{\hbox{\rm Im}\Sigma(\nu)}{[\nu
- \hbox{\rm Re}\Sigma(\nu)]^2 +
[\hbox{\rm Im}\Sigma(\nu)]^2},
\hspace*{0.3truecm}
(10d)
$$
where
$\nu$ is the frequency of the incident radiation, $\Sigma(\nu)$ is the
so-called ``memory function'' which contains the dynamics of the
polaron and depends on $\alpha$ and $\nu$.

As an example, in Fig.\,4  the optical absorption spectrum of Fr\"ohlich
polarons for $\alpha=6$ is shown.

It is remarkable that from (10d) in (Devreese, 1972) the three different
kinds
of polaron excitations, studied before only in asymptotic limits, are
seen to appear in the spectra:
\begin{itemize}
\item{a)} scattering states where e.~g. one real phonon is excited (the
structure starting at $\nu=1$);
\item{b)} R. E. S. (Kartheuser {\it et al.}, 1969; Evrard, 1965);
\item{c)} F. C. states (Kartheuser {\it et al.}, 1969; Evrard, 1965).
\end{itemize}
Experimentally only the ``scattering states'' have been seen for free
polarons (Finkenrath {\it et al.}, 1969);
the R. E. S. might play a role for bipolarons (Verbist {\it et al.} 1994).
However, the full structure of Eq.\,(10d) has been
revealed through cyclotron resonance measurements for which $\nu -
\hbox{\rm Re}\Sigma(\nu)$ is replaced by
$\nu - \omega_c - \hbox{\rm Re}\Sigma(\nu)$ so that
the resonance conditions can be tuned by changing $\omega_c$.

The weak-coupling limit of Eq.\,(10d) coincides with the results by
Gurevich {\it et al.} (1962),
whereas the structure of the strong-coupling limit confirms the
identification of the internal polaron excitations by
Kartheuser {\it et al.} (1969).

In pioneering experimental studies Brown                                                    and
co-workers [see (Brown, 1963)]
have combined mobility experiments and cyclotron resonance
measurements to clearly demonstrate the polaron effect. From a
theoretical plot of mobility versus band mass in AgBr compared to
experimental mobility
data at a given temperature, they estimate
the band mass. This allows to calculate $\alpha$ and the polaron
mass $m^{*}$. This value of $m^{*}$ can then be compared to the measured
cyclotron mass. The experimental value $m^{*}/m_b=0.27 \pm 0.01$ is
obtained for AgBr at 18 K to be compared to a theoretical value
$m^{*}/m_b=0.27 \pm 0.05$.
\subsection{\ch{Fr\"ohlich polarons in 2D}}
\subsubsection{Introduction}
\ch{Today} electron systems in \ch{reduced dimensions e.~g. in}
two dimensions like in GaAs-AlGaAs or
MOSFETS are of great interest. Also the electron-phonon interaction and
the polaron effect in such systems receive much attention. For one
polaron, confined to two dimensions, but interacting with a 3D phonon
gas, the Fr\"{o}hlich Hamiltonian remains of the form (1a) with the
following modification for the $V_{k}$ (Peeters {\it et al.}, 1986a):
$$
V_{k}= -i\hbar \omega_{LO}
\left(\frac{\sqrt{2} \pi \alpha}{Ak}\right)^{\frac{1}{2}}
\left(\frac{\hbar}{m_b \omega_{LO}}\right)^{\frac{1}{4}},
\hspace*{0.3truecm}
(11)
$$
valid because of the fact that the electron-polarization interaction is of the
standard form $1/r$ in an arbitrary number of space dimensions. A possible
dependence of the electron-phonon interaction on a concrete physical
mechanism of the electron confinement to two dimensions was considered by
Fomin and Smondyrev (1994).
\ch{The self-energy $\Delta E/\hbar
\omega_{LO}$ for a polaron} in $2D$ for $\alpha
\rightarrow 0$ and $\alpha \rightarrow \infty$ were derived by
Xiaoguang {\it et al.} (1985),
Das Sarma and Mason (1985):
$$
\frac{\Delta E}{\hbar \omega_{LO}}
\equiv\frac{E_{2D}(\alpha)}{\hbar \omega_{LO}}
=-\frac{\pi}{2} \alpha - 0.06397 \alpha^{2}
+ {\em O}(\alpha^{3})    \hspace*{0.2truecm}  (\alpha \rightarrow 0),
\hspace*{0.3truecm}
(11a)
$$
$$
\frac{\Delta E}{\hbar \omega_{LO}}
\equiv\frac{E_{2D}(\alpha)}{\hbar \omega_{LO}}
=- 0.4047 \alpha^{2}
+ {\em O}(\alpha^{0})    \hspace*{0.2truecm}  (\alpha \rightarrow \infty).
\hspace*{0.3truecm}
(11b)
$$
\ch{The corresponding results for the polaron mass in 2D are
[Xiaoguang {\it et al.} (1985)]}:
$$
\frac{m^{*}}{m_b}
\equiv\frac{m^*_{2D}(\alpha)}{m_{2D}}
=\frac{\pi}{8} \alpha + 0.1272348 \alpha^{2} + {\em
O}(\alpha^{3})  \hspace*{0.2truecm} (\alpha \rightarrow 0),
\hspace*{0.3truecm}
(11c)
$$
$$
\frac{m^{*}}{m_b}
\equiv\frac{m^*_{2D}(\alpha)}{m_{2D}}
=0.733 \alpha^{4} + {\em
O}(\alpha^{2})  \hspace*{0.2truecm} (\alpha \rightarrow \infty).
\hspace*{0.3truecm}
(11d)
$$
The result $\Delta E /\hbar \omega_{LO}= -(\pi/2) \alpha$ was \ch{first} obtained
by Sak (1972).
\subsubsection{Scaling relations}
Peeters and Devreese (1987) have derived several
scaling relations connecting the \ch{polaron} self-energy, the effective mass, the
impedance $Z$ and the mobility $\mu$ in $2D$ to the same quantities in
$3D$. Those relations were derived \ch{on the level of} the Feynman
approximation and are listed here:
$$
E_{2D}(\alpha) = \frac{2}{3} E_{3D}\left(\frac{3\pi}{4}\alpha\right),
\hspace*{0.3truecm}
(12a)
$$
$$
\frac{m^{*}_{2D}(\alpha)}{m_{2D}}=\frac{m^{*}_{3D}\left(\frac{3\pi}
{4}\alpha\right)}{m_{3D}},
\hspace*{0.3truecm}
(12b)
$$
$$
Z_{2D}(\alpha, \nu)= Z_{3D}\left(\frac{3\pi}{4}\alpha, \nu \right),
\hspace*{0.3truecm}
(12c)
$$
where $\nu$ is the frequency of the external electromagnetic field, and
$$
\mu_{2D}(\alpha)=  \mu_{3D}\left(\frac{3\pi}{4}\alpha\right).
\hspace*{0.3truecm}
(12d)
$$
Except for the investigations
on polarons at the surface of liquid He (Jackson, Platzman, 1981;
Devreese, Peeters, 1987), the experimental
studies performed at present, are related to systems with weak
electron-phonon coupling. Those studies have addressed a variety of
physical properties including magneto-phonon anomalies, optical
absorption, plasmon-LO-phonon mode coupling, cyclotron resonance,
mobility, many-body effects, etc. \ldots \ch{The reader is referred
to (Devreese, Peeters, 1987) for more details and additional references.
In what follows some of these studies will be further discussed.}
\subsection{\ch{Fr\"ohlich Polarons in a magnetic field}}
\subsubsection{In 3D}
\ch{Fr\"ohlich polarons have been most clearly manifested by
investigations of their properties in magnetic fields. Therefore a
special section is devoted to the study of polarons in magnetic fields.}
\paragraph{\ch{1.4.1.1. Level crossing, pinning.}}
In interpreting the (low field) cyclotron resonance experiment of Brown
{\em et al.} it was supposed that the zero-magnetic-field polaron mass
is observed.

Of course there exists no {\it a priori} guarantee that this supposition is
true and theoretical studies of polarons in magnetic fields are
necessary. Larsen (1972, 1991)
has made important contributions
to the theoretical study of  polarons in magnetic fields. In
particular he was the first to point out the level repulsion close to
the crossing of levels at $\omega_{c}=\omega_{LO}$ ($\omega_{c}$ is the
cyclotron resonance frequency) and the pinning of Landau levels to the
phonon continuum as $\omega_{c} \rightarrow \infty$ (see Fig.\,5).
Measurements on InSb (Johnson, Larsen, 1966) and CdTe
(Waldman {\it et al.}, 1969) \ch{provided first indications}
for these level crossing and pinning phenomena. Detailed lineshape
studies \ch{for weak coupling} of the cyclotron resonance,
revealing a double peak structure
close to $\omega_{c}=\omega_{LO}$, are displayed by Vigneron {\it et al.}
(1978) and
independently by Van Royen and Devreese (1981).
\paragraph{\ch{1.4.1.2. Static and dynamic properties of polarons in a
magnetic field.}}
Peeters and Devreese (1982) have generalized the Feynman
model of the polaron to the case where a static external magnetic field
is applied. The calculation is valid for all $\alpha$, $\omega_{c}$ and
temperature described by the parameter $\beta$.
The starting point is the
expression of the free energy of the polaron as a path integral
$$
F=F_{ph}
-\frac{1}{\beta} \ln \left\{  \int d{\bf r}
\int_{{\bf r}(0)={\bf
r}}^{{\bf r}(\beta)={\bf r}}   {\cal D}{\bf r} (u) \exp(S[r(u)]) \right\}.
\hspace*{0.3truecm}
(13a)
$$
The ``electron'' contribution to the action $S$ is
$$
S_{e}=-\frac{1}{2} \int_{0}^{\beta} du [{\bf \dot r}(u)^{2}
+i\omega_{c} (x(u)\dot y(u) -y(u) \dot x(u))],
\hspace*{0.3truecm}
(13b)
$$
where
${\bf r}$  is the position vector of the electron, with components $x$,
$y$ in the plane perpendicular to the magnetic field.

A quadratic, retarded model interaction was introduced
(Peeters, Devreese, 1982)
to simulate the polaron (retarded Coulomb)
interaction and, in analogy to the zero-magnetic field case, the
Jensen-Feynman inequality was used:
$$
F \leq F_{ph} + F_{m} -\frac{1}{\beta} <S-S_{m}>_{m}.
\hspace*{0.3truecm}
(13c)
$$
Here $F_{ph}$, $F_{m}$ stand for the free energy of the phonon bath and the
quadratic model respectively, while $<\cdot>_{m}$ denotes an evaluation
of the corresponding average with
$$
{\mbox {\rm e}}^{S_{m}}/\int d{\bf r}
\int_{{\bf r}(0)={\bf r}}^{{\bf r}(\beta)={\bf r}}
{\cal D}{\bf r}(u) {\mbox {\rm e}}^{S_{m}}
\hspace*{0.3truecm}
(13d)
$$
as weight factor. For details the reader is referred to
(Peeters, Devreese, 1982),
where results for the energy and related properties of
the polaron in a magnetic field for all $\alpha$, $\omega_{c}$, $\beta$
were derived both numerically and --- in a variety of limiting
cases --- analytically.

A question of considerable significance raised about the validity of
the inequality (13c) in the presence of a magnetic field. Feynman
suggested that in a magnetic field this inequality remains valid,
or might need a slight modification only (Feynman, Hibbs, 1965).
Several works have been devoted to this challenging problem, see
(Brosens, Devreese, 1988), Sec.\,3.4 in (Fomin, Pokatilov, 1988), and
(Larsen, 1991) for a detailed review.
Larsen (1985) revealed that the groundstate levels of a 2D-polaron obtained
variationally on the basis of (13c) for sufficiently high magnetic fields
lie below those found within the framework of the adiabatic strong-coupling
theory or the fourth-order perturbation weak-coupling approach. Interpreting
the latter groundstate level as the exact one, Larsen came to the conclusion
that it would seem difficult to attach any particular sense to the variational
groundstate level. It is worthwhile that the perturbation groundstate
energy can however itself be treated in some cases [see (Fomin, Pokatilov,
1988), {\it loc. cit.}] as an upper bound for the exact energy. But
when comparing various upper bounds for the exact energy, one should recall
Feynman's (1955) warning that attempts to improve an upper bound by
calculating the higher-order correction terms may indeed deprive the treatment
of its variational nature! Using a model calculation, Brosens and Devreese
(1988) rigorously demonstrated that for sufficiently small electron-phonon
coupling the presence of a magnetic field is prohibitive for the application
of the Jensen-Feynman inequality.

A generalization of the Jensen-Feynman inequality, which remains valid in
the case of a nonzero magnetic field, was derived by Devreese and Brosens
(1992) starting from the ordered operator calculus.
On these grounds, the conditions were determined to be
imposed on the variational parameters in the model action $S_m$, such that
the Feynman upper bound in its original form of the inequality (13c)
remains valid for a polaron in a magnetic field.
Although sofar it has not been conclusively established that a choice of the
parameters in the trial action made in (Peeters, Devreese, 1982) limits them
to the domain determined by the conditions derived in the above-cited work,
it is interesting to note that most of the existing theories of polarons in
a magnetic field (Hellwarth, Platzman, 1962;
Marshall, Chawla, 1970; Evrard {\it et al.}, 1970; L\'epine, Matz,
1976) are obtained as special cases of the results by
Peeters and Devreese (1982) who simply accepted the inequality (13c)
as the working hypothesis.
A prediction by Peeters and Devreese (1982) is that
some quantities characterizing the internal structure of the polaron
(called e.~g. $(v_{\perp}/w_{\perp})^{2}$ in the mentioned paper)
undergo a
drastic change for a well-defined magnetic field (``stripping
transition''). Although high magnetic fields are needed (e.~g.,
$\sim$ 42 T in AgBr), it would be interesting to investigate this
point experimentally.

\paragraph{1.4.1.3. Cyclotron resonance spectra.}
The observation of the cyclotron mass of electrons in AgBr in the low
field case ($\omega_{c} \rightarrow 0$) gave evidence for the occurrence
of polaron effects. However it concerns only one number ($m^{*}/m_{e}$) and
involves the combination of two measurements ($m_{e}$ is the mass of the
electron in vacuum).

It would therefore be useful to analyze the magnetic field dependence of
the polaron mass in order to gain quantitative insight into the validity
of the polaron picture.

An excellent occasion to realize such an analysis is provided by the,
more recent, precise cyclotron mass measurements (in AgBr and AgCl)
by Hodby {\em et al.} (1987). These measurements, performed with
the $5$ cm bore hybrid magnet at Oxford,
cover the range from zero magnetic field to $16$ T.
These measurements are precise enough to distinguish between
various polaron theories.
Several theories were compared in analyzing the experimental data of
by Hodby {\it et al.} (1987).

First, the variational calculation of Larsen (1972) was considered.
This approach is a so-called intermediate-coupling theory to calculate
the energy levels (modified Landau levels) of a polaron in a magnetic field.
The polaron mass is then defined from the energy differences between the
polaron (Landau-) energy levels.

In
principle, it is better to calculate the magneto-optical absorption
spectrum of the polaron (the quantity which is \ch{actually} measured)
and to define
the polaron mass, in the same way as the experimentalist, from the peak
positions in the spectrum. Starting from the results of Peeters and Devreese
(1982), the magneto-optical
absorption of polarons for all $\alpha$ and $\omega_{c}$ at $T=0$
\ch{was calculated} by Peeters and Devreese (1986). They
\ch{evaluated} the memory-function formalism to generalize the study of the
response of a Feynman polaron in a magnetic field. The
magneto-absorption is then obtained from
$$
\lim_{\varepsilon \rightarrow 0} \frac{1}{\nu -\omega_{c} -\sum (\nu+i
\varepsilon)},
\hspace*{0.3truecm}
(14a)
$$
where $\nu$ is the frequency of the incident radiation and $\sum(\nu
+i\varepsilon)$ is the memory-function. The key ingredient of $\sum(\nu
+ i\varepsilon)$ is the space Fourier transform of the density-density
correlation function:
$$
<e^{i{\bf k} {\bf \cdot} {\bf r}(t)}e^{-i{\bf k}{\bf \cdot}{\bf r}(0)}>,
\hspace*{0.3truecm}
(14b)
$$
where $<{\bf \cdot}>$ can be expressed as a path integral. $\sum(\nu +
i\varepsilon)$ is \ch{an intricate} function which takes into account
all the polaron internal states and all the Landau levels.
It turns out that
the magneto-absorption calculated by Peeters and Devreese (1986) leads to the best
quantitative agreement between theory and experiment
as was analyzed for AgBr and AgCl (Hodby {\it et al.}, 1987). It should be
pointed out that the weak-coupling theories (Rayleigh-Schr\"{o}dinger
perturbation theory, Wigner-Brillouin one and its improvements) fail (and
are all off by at least $20 \%$ at $16$ T) to describe the
experimental data for the silver halides.
The analysis of Hodby {\it et al.} (1987) provides
a confirmation of the Fr\"{o}hlich description of the polaron in a case
where weak-coupling approximations are adequate.

An interesting case is provided by the cyclotron resonance data for
CdTe (Johnson, Larsen, 1966). The early analysis (Larsen, 1972) of these
experiments as well as subsequent cyclotron-resonance measurements of the
polaron effective mass in $n$-CdTe (Litton {\it et al.}, 1976)
seemed to suggest that the Fr\"ohlich coupling constant as large as
$\alpha =0.4$ could explain the
data for the polaron mass as a function of magnetic field. This gave rise to
a long-standing challenge to harness the adequacy of the dielectric continuum
model for the magneto-bound polarons in CdTe which maintains the value
$\alpha \sim 0.3$ of the coupling constant.
But the experimental data on the Zeeman splitting in the $1s \to 2p$
shallow-donor-impurity transitions in CdTe at high magnetic
fields  obtained by Cohn {\it et al.}, (1972) have been precisely described
in the framework of a second-order perturbation theory
with band non-parabolicity taken into account using the value
$\alpha = 0.286$ for the Fr\"ohlich coupling constant following from
Eq.\,(1c) (Shi {\it et al.}, 1995). It has been revealed in the latter work,
that Cohn {\it et al.} (1972) had to use as a fitting parameter the value
$\alpha =0.4$ higher than the above-mentioned Fr\"ohlich coupling constant,
in order to compensate the underestimation of the polaron effects in the
calculation of the transition energies.

\subsubsection{Cyclotron resonance of polarons in 2D}
Cyclotron resonance experiments have been performed on the 2DEG, e.~g., in
InSb  inversion layers  and in GaAs-Al$_{x}$Ga$_{1-x}$As
heterostructures (Scholz {\it et al.}, 1983; Seidenbuch {\it et al.}, 1984;
Sigg {\it et al.}, 1985; Merkt, 1985).

Several theoretical studies  have been presented to analyse these
experimental results \ch{for the 2DEG}. In those works the cyclotron mass was obtained
from the positions of the energy levels (Das Sarma, 1984; Larsen, 1984a,b;
Peeters, Devreese, 1985; Peeters {\it et al.}, 1986b,c).

The theory of cyclotron resonance in the 2DEG, for
cases where the electron-phonon interaction plays a significant role
\ch{is reviewed by Devreese and Peeters (1987)}.
Like in 3D (Peeters, Devreese, 1986)
the theory \ch{is expressed} in terms of the memory
function formalism. In \ch{this} treatment the magneto-optical absorption
itself is calculated and the transition frequencies (rather than the
individual energy levels) are obtained directly. \ch{Here} the application
was limited to the weak-coupling regime.

\ch{Some results of this work are}:
\begin{itemize}
\item{a)}
first the magneto-absorption spectrum was calculated, at weak coupling,
for one polaron in 2D. A Landau-level broadening parameter is introduced
phenomenologically in order to remove the divergencies in the
magneto-optical absorption spectrum. The effect of the nonzero width of
the 2DEG is incorporated along with nonparabolicity.
\ch{The} experimental data for $p$-InSb inversion layers can be adequately
explained by this theory.
\item{b)}
To account for the cyclotron-mass data in GaAs-Al$_{x}$Ga$_{1-x}$As
heterostructures it is essential to include many-body effects. Both the
``occupation effect'' (Pauli principle) and the effects of screening were
included, on top of the effect included in the one-polaron studies under
a).
\end{itemize}

It is also worth mentioning that, although for one electron the polaron
effect is enhanced by the 2D-confinement, in reality e.~g., in
GaAs-Al$_{x}$Ga$_{1-x}$As with $n_{e} \geq 1.4 \times 10^{11}$  $cm^{-2}$
screening helps to reduce the polaron effect in 2D so that it becomes
smaller than its 3D counterpart. \ch{For} sufficiently large densities the polaron mass in 2D is not a
monotonically increasing function of $\omega_{c}$ but shows structure
where the filling factor becomes an integer (Peeters {\it et al.},
1988a,b).

It is obvious, both in the case of InSb and that of
GaAs-Al$_{x}$Ga$_{1-x}$As that polaron effects do occur, even if these are
weak-coupling materials.

\ch{A nice example, clearly demonstrating the polaron coupling is
provided by the cyclotron resonance in a 2DEG which naturally occurs in
InSe where $\alpha\approx 0.3$ (Nicholas {\it et al.}, 1992).
One clearly sees, over a wide range
of magnetic fields, the two distinct polaron branches separated by as
much as $0.4\omega_{LO}$ at resonance (Fig.\,5).
Polaron cyclotron resonance has even
been observed in n-type ZnS up to 220 T by Miura {\it et al.}
(1994)}.

A quantitative interpretation of the cyclotron resonance measurements
in $n$-GaAs and AlGaAs-GaAs heterojunctions was obtained on the
grounds of the polaron theory with taking into account three factors:
dimensionality, band nonparabolicity and screening (Sigg {\it et al.},
1985).
Impurity-bound resonant magnetopolarons have been clearly observed in bulk
GaAs and GaAs-Al$_x$Ga$_{1-x}$As multiple quantum wells (Cheng {\it et al.},
1993).

\subsubsection{Formal developments}
Also to treat Fr\"ohlich polarons in a magnetic field the Feynman path
integral proved to be a most powerful method extending the possibilities
of perturbation theory and Hamiltonian variational calculations.

From the methodological point of view the introduction of the total
angular momentum operator of the polaron
$$
\hat{L}_z = {\hat l}_z + i\hbar\sum_{\bf k, \bf k'} a^{\dagger}_{{\bf k}'} a_{\bf k}
(k_x\frac{\partial}{\partial k_y} - k_y\frac{\partial}{\partial k_x})
\delta_{{\bf k},{\bf k}'}
\hspace*{0.3truecm}
(15)
$$
(with ${\hat l}_z$ denoting the electron angular momentum)
is useful for
Hamiltonian treatments of polarons in a magnetic field (Evrard {\it et al.},
1970).
Later this operator was used in studies on excitons,
An operator algebra method was developed by Larsen (1984), useful to study
higher-order effects for polarons in magnetic fields, when expanding in
powers of $\alpha$.

\subsection{The bound polaron}
A polaron can be bound to a charged vacancy or to a charged
interstitial. To a first approximation this system can be approximated
by adding the Coulomb potential energy operator
($-e^2/\epsilon_\infty|\bf r|\equiv -{\tilde \beta}/|{\bf r}|$, $\bf r$ is
the vector operator
characterizing the electron position with respect to the center of the
vacancy or of the interstitial) to the Fr\"ohlich Hamiltonian.

Intuitively one expects that the weak-coupling polaron spectrum for the
bound polaron is approximately given by a Bohr formula adapted to take
into account the polaron mass:
$$
E_n = -\alpha\hbar\omega_{LO} - \frac{m_b e^4}{2\hbar^2n^2\epsilon_0^2}
- \frac{\alpha}{12}\frac{m_b e^4}{\hbar^2 n^2} + O(\alpha^2)\>
(n=1,2,\ldots).
\hspace*{0.3truecm}
(16)
$$

Expansions refining Eq.\,(16) were derived
first using approximate schemes for
perturbation theory (Bajaj, 1972); later a rigorous result for the binding energy
$E_n$ up to order $\alpha$ was obtained (Engineer, Tzoar, 1972).

Further schemes to treat  the groundstate energy of the bound polaron
(and approximations for some of the excited states) have been developed
by Platzman (1962), Larsen (1969), Devreese {\it et al.} (1982), Adamowski
(1985).

Brandt and Brown (1969) have interpreted some structure in their infrared
optical absorption spectra of AgBr as caused by the bound polaron; in
particular the 168 cm$^{-1}$ absorption line has been analyzed as a
transition between a 1$s$ and a 2$p$ state (modified by the polaron
interaction). Also higher excited states and LO-phonon sidebands play a
role in this spectrum [see e.~g. (Bajaj, 1972)] (Fig.\,6).
The bound polaron is related to the F-center. In Tables 2 and 3 some
energy levels of the bound polaron are tabulated.
\section{The Small Polaron}    \label{secsp}

\subsection{The small-polaron concept. Role of localization}

An electron or a hole trapped by its self-induced atomic (ionic)
displacement field in a region of linear dimension (``radius''), which
is of the order of the lattice constant, is called {\it small polaron}
(Fr\"ohlich, 1957; Sewell, 1958; Fr\"ohlich, Sewell, 1959;
Holstein, 1959; Emin, Holstein, 1976). An excellent survey of the
small-polaron physics relevant to the conduction phenomena in non-crystalline
materials and to the metal-insulator transitions has been given
by Mott (1987, 1990).
As distinct from large
polarons, small polarons appear due to short-range forces. Thus,
in \ch{certain} materials, in particular in some oxydes, the induced lattice
polarization is essentially localized in a volume of the order of a unit
cell.
Hence, the charge carrier is localized on an individual lattice site during a
time \ch{which can become} large compared to the {\it localization time}
describing the relaxation of the lattice to the small-polaron state.
The localization time is of the same order of magnitude as the period of
a lattice vibration, $\omega_{LO}^{-1}$.

Because for small polarons the lattice polarization is mostly confined
to one unit cell, the atomicity of the solid is felt by the carrier; a
complete treatment of small polarons should therefore start from an {\it ab
initio} calculation which takes into account the detailed local structure
of the solid; the Fr\"ohlich continuum approximation would not be
adequate. Nevertheless, actual small-polaron theories as developed, e.~g.,
by Yamashita and Kurosawa (1958), Holstein (1959) and others are based on
analytical approximations \ch{as a starting point}.
Thus, the adiabatic eigenstates of an electron placed in a deformable
continuum were shown to depend drastically on the character of the
electron-lattice interaction as well as on the dimensionality of the system
(Emin, Holstein, 1976). As distinct from the case of the long-range
interaction with a stable large-polaron state, for the short-range
interaction in a three-dimensional system there exist two stable states,
namely, an unbound electron in an undeformed continuum and an electron
collapsed in an infinitesimally localized self-induced potential well.
The former is analogous to the band electron state in a rigid lattice, the
latter models a small-polaron state. It is also worthwhile that the
common action of a long-range and of a short-range forces was found to yield
always a small-polaron-like state and --- in a certain region of the
interaction strengths --- a large-polaron state. Thus, even from the early
analysis a possibility of a {\it coexistence} of the both types of
polarons can be distinctly deduced.

In the modern theories [see, e.~g. (Alexandrov, Mott, 1994)],
the small-polaron energy is regarded to consist of the following parts:
\begin{itemize}
\item{a)} the kinetic energy of the charge carrier in a rigid lattice which,
as distinct from large polarons, is considered to originate from the
intersite transfer due to tunneling;
\item{b)} the energy of the atomic (ionic) displacements field, describing
the lattice distortion;
\item{c)} the potential energy of the charge carrier in the potential well
formed by these displacements.
\end{itemize}

The interaction of the localized electron (hole) with the lattice
vibrations then induces the charge carrier to jump from one atom (or
ion) to a neighbouring one. This process is called {\it hopping}.
The detailed physical picture of
hopping (Lang, Firsov, 1962) suggests a sequence of the acts of small-polaron
disintegration and reappearance as follows.
At sufficiently high temperatures $k_BT>\hbar\omega_{LO}/4$, the typical
time interval between jumps $\Delta t$ satisfies the inequalities:
$t_0 \ll \Delta t \ll t_p$, where
$t_0 \sim \hbar/[(W_H k_B T)^{1/2}]$
(with $W_H$, the thermal activation energy for hopping)
denotes the {\it jump-over} time and
$t_p \sim \hbar/ \Delta E_p$
(with $\Delta E_p$, the small-polaron bandwidth)
is the tunneling time. Hence, an electron remains most of the time
at a site, suffering a hopping transition from site to site rather rarely, but
on average earlier, than a tunneling occurs.
As far as the jump-over time is much shorter than the period of a
lattice vibration, $t_0 \ll \omega_{LO}^{-1}$, under a hopping transition
the electron ``jumps out of'' the old potential well due to its self-induced
lattice deformation, thus initiating a multiphonon process of the lattice
relaxation: the small-polaron state disappears.
But the time interval between jumps is much larger than the
localization time, $\Delta t \gg \omega_{LO}^{-1}$. This inequality describes
the {\it anti-adiabatic limit}:
the atoms (ions) can adiabatically follow the
motion of an electron, contrary to the case of a large polaron
(see section 1.2.1). Thus, a new potential well due to the lattice
deformation adapted to the new position of the electron is formed:
a renascence of a small-polaron state occurs.

Also in the case of small polarons the relevant phonons are commonly the
LO phonons. A simple estimate of the temperature dependence of mobility
can be obtained starting from the following The larger the number of LO phonons
$n_{ph}$ present in the solid, the larger the mobility \ch{$\mu_{SP}$
of the small polaron}:
$$
\mu_{SP} \sim n_{ph}.
\hspace*{0.3truecm}
(17)
$$
For sufficiently low temperatures \ch{$n_{ph} =
e^{-\hbar\omega_{LO}/kT}$} and, as a consequence, we find
$$
\mu_{SP} \sim    e^{-\hbar\omega_{LO}/kT}.
\hspace*{0.3truecm}
(18)
$$
It should be emphasized that the mobility of small polarons is therefore
{\em thermally activated} and its temperature dependence is totally
different from that of Fr\"ohlich polarons \ch{[compare e.~g. Eq.\,(18) with
Eq.\,(10a)]}.

A more detailed theoretical treatment of the small-polaron mobility
[see (Lang, Firsov, 1962; Reik, 1972)]
leads to the following formula, valid for  $T>\theta_D/2$ ($\theta_D$ is
the Debye temperature of the crystal):
$$
\mu_{SP} = \frac{ea^2\omega_{LO}}{6k_BT} \exp\left(-\frac{W_H}{k_BT}\right),
\hspace*{0.3truecm}
(19)
$$
where $a$ is the lattice constant of the crystal in which the small polaron
occurs, $W_H$ is the thermal activation energy for hopping and is given
by 1/2 the small-polaron binding energy.

The Arrhenius-type activated behaviour of the form (19) of a mobility
has been used as a
fingerprint to identify small-polaron behaviour in solids. One of the
earliest studies concerned one of the  uranium oxydes  UO$_{2+x}$
(Devreese, 1963; Nagels {\it et al.}, 1964).
Subsequently small polarons were invoked to interpret the conductivity
in  many  oxydes (Mott, Davis, 1979), in particular in transition metal
oxydes. It should
be mentioned that often the measured quantity is the Hall-mobility
rather than the drift mobility. The theory of the Hall mobility of small
polarons due to hopping (Friedman, Holstein, 1963; Austin, Mott, 1969)
leads to
$$
\mu_{Hall} \propto T^{-{1\over 2}} \exp\left(-{W_H \over 3k_BT}\right).
\hspace*{0.3truecm}
(20)
$$
The relation between Hall and drift mobility is not simple, and the Hall
mobility depends, e.~g., on the interference between several hopping
processes [see a comprehensive review by Austin and Mott (1969)].
For $s$-carriers the Hall coefficient \ch{turns out to be}
always negative.

\subsection{Standard small-polaron theory}

As shown above, small polarons --- at sufficiently high temperature --- are
characterized by diffusive motion and the band-picture with its Bloch
states breaks down; in the low-temperature limit the band picture
reappears in the theoretical description  although experimental
evidence for \ch{band conduction} has been limited.

The fact that
the Bloch-band picture breaks down is connected with the narrowing
of the band gap which develops as the carrier becomes more and more
localized, resulting in an increasing effective mass and, in the limit,
in self-trapping. \ch{E.~g.} in KCl the material characteristics are
such that a valence band hole gets  self-trapped  due to its polaron
interaction with the lattice (Stoneham, 1979). The self-trapping of the
hole is a subtle process, but the evidence is that it is related to the
polaron formation in interplay with the Jahn-Teller effect (Stoneham, 1979).

For the quantitative treatment of small polarons the so-called molecular
crystal model of Holstein (1959) is perhaps  most illuminating.
Without going into the mathematical details of this model, we
mention its basic ingredients: a linear  (1D) chain is considered
with $N$ diatomic molecules in which an excess electron is moving.
With this model, the occurrence of two regimes, separated by a
characteristic temperature typically of order 0.4 to 0.5
$\hbar\omega_{LO}/k_B$,  is established theoretically: \ch{a) hopping induced
by phonons and b) Bloch-type band motion}.  For hopping motion of
small polarons, Holstein derived the following expression  for the
hopping mobility:
$$
\mu = \frac{ea^2}{k_BT}\frac{J^2}{\hbar^2\omega_{LO}}
\left[ \frac{\pi}
{\gamma {\rm cosh} \left( \frac{\hbar\omega_{LO} }{4k_B T} \right)}
\right]^{1\over2}
\exp\left[{-2\gamma \tan \left(\frac{ \hbar\omega_{LO} } {4k_B T}
\right)}\right],
\hspace*{0.3truecm}
(21)
$$
where $J$ is a two-center overlap integral, $\gamma$ is a measure for the
electron-phonon coupling strength  for small polarons to be
distinguished from the large polaron coupling constant $\alpha$
($2J$ corresponds to the width of the electronic Bloch band which is supposed
to be relatively small in small-polaron theory).

An important role in small-polaron theory belongs to the distinction between
adiabatic and non-adiabatic hopping transitions [roughly speaking,
the adiabatic regime is characterized by the fact that the electron
follows the atomic (ionic) motion instantaneously], see for the details
the monograph by Klinger (1979).

The formation of small bipolarons by coupling of electrons to acoustic
phonons and in disordered media was examined by Cohen {\it et al.} (1984).
The study of small-polaron properties has been extended and  studied in
depth  by
Mott who identified and analyzed many instances of small-polaron
transport  including variable-range hopping, in which electrons hop over
a range of distances and not only between nearest neighbours, and the
role of small polarons in amorphous semiconductors (Mott, 1990).
The coherence and dynamics of small polarons
in the presence of disorder were represented in terms of two characteristic
energies: the polaron bandwidth specifies the energy scale of
disorder at which the polarons become localized as composite particles, while
the bare electron bandwidth defines the energy scale at which the
polaron ceases to be a composite particle (Spicci {\it et al.}, 1994).

In the theory of small polarons there are still some open fundamental
questions which have been a subject of considerable recent investigation.
Among the urgent issues: (i) the problem of the relevance of the Bloch-like
states for a single small polaron in spite of the retardation, and (ii)
the study of the nature and properties of quasiparticles in many-polaron
systems, which are of especial interest for the polaron,
bipolaron and hybrid polaron-bipolaron pictures of high-T$_C$
superconductivity (see Sec. \ref{HTC}) should be pointed out.

\subsection{Experimental evidence}

Experimentally small-polaron effects have been analyzed, e.~g., in KCl, LiF,
NiO, MnO, TiO$_2$, BaTiO$_3$, SrTiO$_3$, LaCoO$_3$,...
We refer to the reviews by Appel (1968) and Firsov (1975) for more details.
More recently, de Jongh (1988),
Micnas {\it et al.} (1990),
Alexandrov and Krebs (1992) and Alexandrov and Mott (1994)
surveyed in detail both the principles and the main results of the
small-polaron theory in the
context of the (bi)polaronic approach in the physics
of high-T$_C$ superconductors and
tried to interpret some experimental data in different materials
in terms of small polarons and small bipolarons.

The study of the optical absorption for small polarons is complex.
A representative example is shown in \ch{Fig.~7},
where the real part of the ac conductivity, describing the
small-polaron absorption, as derived from the Kubo formula, is shown
for various $\Gamma$ ($\Gamma = \hbar/4\tau_0 k_B T$, where $\tau_0$ is
1/4 times the ``hopping time'') (Reik, Heese, 1967). \ch{Note the completely
different character of the optical absorption for small polarons as
compared to large polarons.}

In analyzing experimental transport data  also  thermoelectric power
measurements are used; the theoretical study of the thermopower for
small polarons has revealed that no polarization energy is transferred by
the polaron motion.

Mobile polarons, observed in WO$_{3-x}$ by Gehlig and Salje (1983), were
shown to exhibit at 130 K a transition from a regime of hopping
conductivity, characterized by a constant activation energy, to a regime
of band conductivity, in which the process is not activated.
With increasing carrier density,
small polarons are formed up to a density, which is equal to the
concentration of the sites at which they can be localized. At this
critical density the dc electrical conductivity shows a phase transition,
which these authors interpret as an Anderson-type transition:
a change from a thermally activated
small-polaron behaviour to a metallic temperature dependence occurs.
the critical density was found to be about 3.7$\times$10$^{21}$ cm$^{-3}$
in WO$_{3-x}$ (Salje, G\"uttler, 1984) and 1.7$\times$10$^{21}$ cm$^{-3}$
in NbO$_{2.5-x}$ (R\"uscher {\it et al.}, 1988). At higher
densities, two type of carriers are suggested to coexist: small polarons, on the
one hand, and, on the other hand,
conducting carriers which can be regarded as large polarons.
The crossover
from  small-polaronic to  metallic temperature dependence of the
conductivity is consistently demonstrated by the Arrhenius plot for five
different chemical compositions NbO$_{2.5-x}$.
A thermally
activated small-polaron conductivity seems to occur at lower degrees of
reduction than that of NbO$_{2.49}$. This crossover
scenario is strongly supported by the fact, that close to the same
critical densities as mentioned above  the saturation of the ``integral
intensity'' of the polaronic absorption occurs, see Fig.\,8.

Quite recently, the dc electrical
conductivity of a slightly hyperstoichiometric sample of polycrystalline
UO$_{2+x}$ was interpreted (Casado {\it et al.}, 1994)
in the framework of small-polaron theory, where some discrepancies between
the semi-empirical values of the small-polaron self-energy and the thermal
activation energy, from the one side, and those obtained as a result of a
fully microscopic calculation, from the other side, are revealed.

The recent measurements of the Seebeck coefficient of
BaBi$_{0.25}$Pb$_{0.75}$O$_{3-\delta}$ with an oxygen deficiency
by Hashimoto {\it et al.} (1995)
suggest the coexistence between the band charge carriers of high mobility, on
the one side, and localized charge carriers of low mobility, on the
other side (Fig.\,9).
Namely, in the low-temperature region, $T < 200$ K, the decrease
of the (negative) Seebeck coefficient with $T$ is supposed to be due
to large electron polarons, while  a temperature-activated behaviour of the
(positive) Seebeck coefficient above room temperature is attributed by
the authors to small hole polarons.
A non-phenomenological interpretation of such experiments requires a
theoretical approach which would combine the large- and the small-polaron
concepts.

A complex of experimental results on dielectric relaxation, ac and dc
conductivities of La$_{1-x}$Sr$_{x}$FeO$_3$ with 0.05 $\leq x \leq$ 0.3
obtained by Jung and Iguchi (1995) was self-consistently explained in
terms of small polarons (see, e.~g., Arrhenius plot for conductivity in
Fig.\,10).

\section{\ch{Bipolarons and Polaronic Excitons}}\label{BIP}
When two electrons (or two holes) interact with each other
simultaneously through the Coulomb force and via the
electron-phonon-electron interaction either two independent polarons can
occur or a bound state of two polarons --- the {\it bipolaron} --- can arise
(Vinetskii, 1961; Hiramoto, Toyozawa, 1985; Adamowski, 1989).
Whether bipolarons originate or not, depends on the competition between
the repulsive forces (direct Coulomb interaction) and the attractive
forces (mediated through the electron-phonon interaction).

The bipolaron can be {\em free} and characterized by translational
invariance, or it can be {\em localized}.
According to Alexandrov and Ranninger (1981 a,b), the
many-electron system on a lattice coupled with any bosonic field turns out
to be a charged Bose-liquid, consisting of small bipolarons in the
strong-coupling regime.

Similarly to the case of a bipolaron, an electron and a hole in a
polarizable medium interacting with each other simultaneously both through
the Coulomb force and via the electron-phonon-hole interaction,
form a quasiparticle --- the {\it polaronic exciton}.

\subsection{Fr\"ohlich  bipolarons}
In this section the case of free bipolarons for electrons \ch{or} holes
interacting with longitudinal optical phonons is discussed \ch{for the
case of the continuum limit}. They are referred to as Fr\"ohlich bipolarons.

Fr\"ohlich bipolarons are described by the following Hamiltonian
\begin{eqnarray}
H &=& \sum_{j=1,2} \left[ \frac{{\bf p}_j^2}{2m_b} + \sum_{\bf k}(V_ka_{\bf k}e^{i{\bf k}
\cdot {\bf r}_j} + V^*_ka^\dagger_{\bf k}e^{-i{\bf k}\cdot{\bf r}_j})\right]
\nonumber\\
&&+\sum_{\bf k} \hbar\omega_k a_{\bf k}^\dagger a_{\bf k}
+ U({\bf r}_1-{\bf r}_2),
\hspace*{0.3truecm}
(22)\nonumber
\end{eqnarray}
where
${\bf p}_j, {\bf r}_j$ characterize the $j$'th electron ($j$=1,2),
the potential energy for the Coulomb repulsion equals
$$
U({\bf r}) = \frac{e^2}{\varepsilon_\infty |{\bf r}|}
\equiv\frac{U}{|{\bf r}|},
\hspace*{0.3truecm}
(23)
$$
$\varepsilon_\infty$ is the high-frequency dielectric constant and the
other symbols in Eq.\,(22) are the same as those in Eq.\,(1a). Note that one
always has
$$
U > \sqrt{2}\alpha
\hspace*{0.3truecm}
(24)
$$
$(\hbar=\omega_{LO}=m_b=1)$: this inequality expresses
the obvious fact that $\varepsilon_0>\varepsilon_\infty$.

In the discussion of bipolarons often the  ratio
$$
\eta = \frac{\varepsilon_\infty}{\varepsilon_0}
\hspace*{0.3truecm}
(25)
$$
of \ch{the} electronic and static dielectric constant is used ($0\le\eta\le 1$).
It \ch{turns} out that bipolaron formation is favoured by smaller
$\eta$. To estimate the order of magnitude of the quantities involved
one may express $\alpha$ and $U$ as follows:
$$
\alpha = \sqrt{\lambda}\left( \frac{1}{\varepsilon_\infty} -
\frac{1}{\varepsilon_0}\right),
\hspace*{0.3truecm}
(26a)
$$
$$
U = \sqrt{2\lambda}\frac{1}{\varepsilon_\infty}
\hspace*{0.3truecm}
(26b)
$$
with
$$
\lambda = \frac{m_b}{m_e}\frac{R^*_y}{\hbar\omega_{LO}}.
\hspace*{0.3truecm}
(26c)
$$
The effective Rydberg is characterized by the electron (hole) band mass $m$:
$$
R^*_y = \frac{m_b e^4}{2\hbar^2}.
\hspace*{0.3truecm}
(27)
$$
Verbist {\it et al.} (1990, 1991) analyzed
the Fr\"ohlich bipolaron using the Feynman path
integral formalism. Quite analogously to the above discussed relations
(12 a to d), a scaling relation was derived between the free
energies $F$ in two dimensions $F_{2D}(\alpha, U, \beta)$ and in three
dimensions $F_{3D}(\alpha, U, \beta)$:
$$
F_{2D}(\alpha, U, \beta) =
\frac{2}{3} F_{3D}(\frac{3\pi}{4}\alpha, \frac{3\pi}{4}U, \beta).
\hspace*{0.3truecm}
(28)
$$
This is the generalization to bipolarons of the scaling relation for a
single Fr\"ohlich polaron \ch{(Peeters, Devreese, 1987)}.
Physically the scaling relation implies that bipolaron formation will be
facilitated in 2D as compared to 3D. (The critical value for bipolaron
formation $\alpha_c$ will be scaled with a factor $3\pi/4\approx 2.36$
or: $\alpha^{(2D)}_c = \alpha^{(3D)}_c/2.36$).

Smondyrev {\it et al.} (1995) derived analytical strong-coupling asymptotic
expansion in inverse powers of the electron-phonon coupling constant
for the large bipolaron energy at $T=0$
$$
E_{3D}(\alpha, u) =
\frac{2\alpha^2}{3\pi} A(u) - B(u) + O(\alpha^{-2}),
\hspace*{0.3truecm}
(29a)
$$
where the coefficients are closed analytical functions of the ratio
$u=U/\alpha$:
$$
A(u) = 4 -2\sqrt{2}u\left(1+\frac{u^2}{128}\right)^{3/2} + \frac{5}{8}u^2-
\frac{u^4}{512}
\hspace*{0.3truecm}
(29b)
$$
and for $B(u)$ see the above-cited paper. The scaling relation (28) allows
to find the bipolaron energy in two dimensions as
$$
E_{2D}(\alpha, u) =
\frac{2}{3}
E_{3D}(\frac{3\pi}{4}\alpha, u).
\hspace*{0.3truecm}
(30)
$$

A ``phase-diagram'' for the polaron---bipolaron system was
introduced by Verbist {\it et al.} (1990, 1991).
It is based on the generalized trial action. This
phase diagram is shown in \ch{Fig.\,11}  for the 3D-case.
A Fr\"ohlich coupling
constant as high as 6.8 is needed to allow for bipolaron formation. No
definite experimental evidence has been provided for the existence of
materials with such high Fr\"ohlich coupling constant. (One of the
highest $\alpha$'s reported is for RbCl where
$\alpha \approx 3.8$ and for CsI where $\alpha\approx 3.7$, see Table 1).

Materials with  sufficiently  large $\alpha$  for Fr\"ohlich bipolaron
formation in 3D might  exist but careful analysis
(involving e.~g. the study of cyclotron resonance), like the one executed
for AgBr, AgCl (Hodby {\it et al.}, 1987) is in order to confirm this.
Presumably some
modifications to the Fr\"ohlich Hamiltonian are also necessary to
describe such high coupling because of the more localized character of
the carriers in this case which makes the continuum approximation less valid.

The confinement of the bipolaron in 2D facilitates bipolaron
formation at smaller $\alpha$. From \ch{Fig.\,12}  it is seen that bipolarons
can now be stable for $\alpha\ge 2.9$, a domain of coupling constants
which is  definitely  realized in several solids. Intuitive arguments
suggesting that
bipolarons are stabilized in going from 3D to 2D  had  been given before
but the quantitative analysis based on the path integral was presented
by Verbist {\it et al.} (1990, 1991).

The stability of bipolarons has also been examined with the use of
operator techniques where the center of gravity motion of the bipolaron
was approximately separated from the relative electron (hole) motion
[see (Bassani {\it et al.}, 1991)]. The results by Bassani
{\it et al.} (1991) and by Verbist {\it et al.} (1990, 1991)
tend to confirm each other.

The bipolaron was also approached (Hiramoto, Toyozawa, 1985)
in the path-integral representation using a special case of the trial
action of (Verbist {\it et al.}, 1990; 1991);
in this work the combined effect of LO phonons, acoustic
phonons and deformation potential was analyzed.

Early work on bipolarons had been based on strong-coupling theory in
which case the bipolaron stability can be expressed with $\eta$ as the
\ch{sole} parameter. ($\eta<0.08$ is a typical strong-coupling result
for bipolaron stability.)
It turns out that  the numerical
stability criteria (Verbist {\it et al.}, 1990)
can be adequately formulated
analytically for all $\alpha$, for which the bipolaron is stable.

A very clear representation of experimental evidences for
bipolarons, e. g. from the data on magnetization and electric conductivity
in Ti$_4$O$_7$, as well as in Na$_{0.3}$V$_2$O$_5$ and polyacetylene, given
by Mott (1990) is to be mentioned.

\subsection{Polaronic excitons}

Excitons constitute very interesting physical entities
which in polarizable media are relevant to polarons:
in polar systems they can be
conceived as two interacting polarons of opposite charges
(Haken, 1956; Toyozawa, 1963, 1964; Knox, 1963;
Bassani, Baldereschi, 1973; Bassani, Pastori Parravicini, 1975;
Adamowski {\it et al.}, 1981; Wallis, Balkanski, 1986).

The exciton groundstate energy in a polar crystal was determined
by Pollman and B\"uttner (1975, 1977) taking into account the fact that
the potential energy of the
electron-hole interaction depends on the quantum state of the interacting
particles due to the polaronic effect. Later on, many works were devoted to
this problem. Similar considerations
were applied by Petelenz and Smith (1981) to explain the
dependence of the binding energy of an exciton-ionized donor complex on the
electron-to-hole mass ratio in CdS and TlCl, and by Larsen (1981)
to show that the ratio of the binding energy of the
$D^-$-centers to that of neutral donors
in AgBr and AgCl is as much as one order of magnitude larger than in
nonpolar crystal. The binding enhancement is due to the {\it attraction}
between the electrons and the static polarization charge induced by them
in the central part of the ion.
The experimental data on spectral photoconductivity in the systems
Ca-Sr-Bi-Cu-O (Masumi {\it et al.}, 1988a) and Ba-Pb-Bi-O
(Masumi {\it et al}, 1988b)
have been interpreted to be due to an exciton-mediated bipolaronic mechanism.
The recent experimental data on
the reflectivity and its temperature dependence in La$_2$CuO$_4$ were
interpreted in terms of polaronic excitons by Falk {\it et al.} (1992).

\subsection{Localized bipolarons}

Localized bipolarons tend to be {\it small} bipolarons, characterized by a
radius of the order of the lattice constant.

For a small bipolaron, both constituting polarons can be
localized either at the same lattice site [intrasite, or Anderson, bipolaron, see
(Anderson, 1975)] or at two different lattice sites, e.~g.
at two neighbouring lattice sites (intersite bipolaron).


For two electrons (holes) on the same site the \ch{direct} Coulomb repulsion
is governed by the potential energy $U$. Whether or not two electrons
remain  at  the same site, is  determined  by $U_{eff}$, the effective
potential which arises if both the Coulomb repulsion and the electron
(hole)-phonon or polaron interaction are taken into account. If the
polaron interaction dominates ($U_{eff}<0$) the Coulomb repulsion one
speaks of {\em negative-U behaviour} and two carriers can be occupying
the same site (``double occupancy of a lattice site''); a bipolaron
localized on one site arises.
Negative-$U$ bipolarons have been suggested to occur e.~g. in {\em
chalcogenide glasses} (Anderson, 1975;  Mott, 1990).

The intersite bipolaron can form {\em singlet} (bonding
as well as anti-bonding) or {\em triplet} states (Fig.\,13).
Intersite singlet bipolarons are usually referred to as Heitler-London
bipolarons, see for example (de Jongh, 1988).

Localization of bipolarons was investigated by many authors including
Lannoo {\it et al.} (1959), Hubbard (1964),
Stoneham and Bullough (1971), Anderson (1975), Mott (1990),
to name a few. \ch{An extensive literature exists and a} more
complete discussion is given by Fisher {\it et al.} (1989).
Localized bipolarons have been studied in \ch{numerous oxydes including}
Ti-oxydes, vanadium bronzes, LiNbO$_3$, WO$_3$.

Hybridization between coexisting localized bipolarons and itinerant electrons
leads to a possibility of the coherent motion of bipolarons and their phase
transition into a superfluid state upon lowering the temperature at a certain
critical temperature (Ranninger, 1994), which seems to be relevant to the
phenomena occurring in cuprates, bismuthates and fullerens.

\section{Spin Polarons}
Just like electrons or holes interact with the Bose-field of phonons, they
can,  because of  the exchange interaction, interact with the magnon
field. This gives rise to the formation of the {\em spin polaron} (or
magnetic polaron).
One of the first studies related to spin polarons is \ch{due to de
Gennes (1960).}

Intuitively one can imagine that e.~g. in an
antiferromagnet, the carrier spin polarizes the spins of the surrounding
magnetic ions. The spin polaron consists of the carrier with its spin
together with the lattice magnetization created by the carrier spin.
Similar to the case of the dielectric polaron the physical properties of
the carrier are influenced by the self induced magnetization cloud. The
mechanism for the formation of spin polarons can be easily understood
for materials with preference for antiferromagnetic alignment; the
kinetic energy of a carrier can be reduced by reversing the spins of the
surrounding ions; this allows the carrier to move more easily around
these ions (Wood, 1991).

In principle spin polarons can arise in ferromagnets as well.
Also the induced magnetic polarization can be
characterized by ferromagnetic as well as antiferromagnetic alignment.

In general it is  accepted  that the evidence for the occurrence of {\em
bound magnetic polarons} is more convincing than that for {\em free
magnetic polarons} (Benoit \'a la Guillaume, 1993).
Bound magnetic polarons occur when the carriers are
localized (e.~g. because they are bound to a donor or to an acceptor).

The bound magnetic polaron has been studied primarily in Mn-based
semimagnetic semiconductors (SMSC). The bound magnetic polaron then
originates from  the  exchange  interaction between a rather localized
unpaired
carrier and the 5/2 spin of the surrounding Mn$^{2+}$ ions. In the bound
magnetic polaron ferromagnetic order occurs within the orbit of the
bound carrier and the magnetic exchange leads to an increase of its
binding energy. For spin polarons, e.~g. in the case of EuS doped with
GdSe (Mott, 1990), the basic interaction energy between a conduction electron spin
and the surrounding magnetic moments is given by
$$
J_{sf}sS|\Psi_s(0)|^2V
\hspace*{0.3truecm}
(31)
$$
$J_{sf}$ is the energy of the ferromagnetic coupling between the spin
$s$ of the conduction electron and $S$ the europium ion spin.
$\Psi_s(0)$ is the \ch{wave function of the conduction electron} at the origin. $V$ is the
atomic volume.

A transparent intuitive consideration due to Mott (1990)
leads to a simple expression for the spin polaron radius:
$$
R = \left(\frac{\hbar^2\pi a^3}{4m^*J_N}\right)^{1/5},
\hspace*{0.3truecm}
(32)
$$
where $J_N$ is the energy, per moment, needed for a transition from the
antiferromagnetic alignment to the  ferromagnetic alignment,
$m^*$ is the effective mass of the carrier,
$a$ the lattice parameter of the host  solid.

For the effective mass of the spin polaron the following expression has
been derived (Mott, 1990):
$$
m^*_{sp} = m^*e^{\gamma R/a}
\hspace*{0.3truecm}
(33)
$$
 with $\gamma\approx 1$.
\section{\ch{Bipolarons and  High-Temperature \\ 
Superconductivity}}\label{HTC}
Dielectric or Fr\"ohlich bipolarons as well as spin bipolarons have been
considered to  possibly  play a role in superconductivity. \ch{Also the
``small-polaron'' interaction was studied in the framework of
superconductivity.}

In fact already the charged bosons of Schafroth (1955)
could be \ch{thought} of as Fr\"ohlich bipolarons. However although the
London equation indeed results from a bipolaron gas \ch{model},
a continuous behaviour of the specific heat $C_V$ was found at the
superconducting transition \ch{temperature} in
contradistinction from experiment. After the success of the BCS theory
with its Cooper pairs consisting of electrons coupling in $\bf k$-space:
$|\!\uparrow,\bf k; \downarrow,-\bf k\rangle$, the bipolaron gas and the
pairing in real space  became  a less central issue.

Nevertheless, theoretical models incorporating small bipolarons
(localized Cooper pairs with electrons (holes) of opposite spin on
nearest neighbour sites) and small bipolarons in narrow bands and
characterized by hopping conductivity were still studied
(Alexandrov, Ranninger, 1981a,b; Chakraver\-ty {\it et al.}, 1987).
These models had not predicted high-$T_C$ superconductivity  however.

After the discovery of high-$T_C$ superconductivity alternative models
(with respect to BCS)
for superconductivity received renewed attention. It is experimentally
clear that also in the high-$T_C$ materials  pairing of carriers takes
place. However different coupling mechanisms might be at the
basis of the pairing; e.~g. coupling through acoustic plasmons  has
been considered.

In the present context  a   preliminary question is whether Fr\"ohlich or
spin bipolarons \ch{do form} in high-$T_C$ materials. This question has
been investigated by Verbist {\it et al.} (1991)
and the polaron-bipolaron phase diagram was
applied e.~g. to La$_2$CuO$_4$ \ch{(see Fig.\,12)}. In  this \ch{figure}  also
YBa$_2$Cu$_3$O$_6$ is \ch{represented}. On the basis of the existing
experimental data it can be stated that the characteristic point  in the
polaron-bipolaron phase diagram   for
La$_2$CuO$_4$ and YBa$_2$Cu$_3$O$_6$ lies on the straight line through
the origin  and shown in the figure. Presumably the Fr\"ohlich
coupling  is large enough  in these materials  that  their
$\alpha$ overlaps with the bipolaron existence region for
$\alpha$. It is remarkable that the existence line for bipolarons
\ch{lies} very close  (La$_2$CuO$_4$) or penetrates (YBa$_2$Cu$_3$O$_6$) the
very narrow existence domain for large bipolarons. It seems therefore
{\em safe to assume} that Fr\"ohlich bipolarons do occur in some of the
high-$T_C$ materials (Verbist {\it et al.}, 1990, 1994).

It should be {\ch re-iterated} that the precise $\alpha$-value characterizing
the high-$T_C$ materials is not known \ch{as long as} the band mass  has not
been determined.
It may also be observed that most other
``polaron materials'' (alkali halides etc.) find their characteristic points
in the polaron-bipolaron phase diagram far away from the bipolaron
stability area.

Cataudella {\it et al.} (1992), Iadonisi {\it et al.} (1995) have undertaken
the study of the
many-body physics of the bipolaron gas in an approximation comparable
\ch{to} that by Verbist {\it et al.} (1990),
but based on a Hamiltonian formulation.
Although the mathematics of this problem \ch{is}
complex and the studies are far from finalized, some promising features
arise; especially the variation of $T_C$ for the high-$T_C$ materials as
a function of the doping is well reproduced, be it with one fitting
parameter.

On the basis of results by Verbist {\it et al.} (1991),
also the optical spectra of bipolarons were analyzed; it was  suggested that
the mid IR-peak characterizing the high-$T_C$ solids is related \ch{to
relaxed excited states} of the bipolaron.

Emin and Hillery (1989) studied the interplay between short- and long-range
polaron interaction in the adiabatic (strong-coupling)
approximation. This work joins the ideas of Schafroth and examines in
more detail the character of the mobile charged bosons as bipolarons.
(It should be noted that  --- in fact --- the adiabatic theory
does not lead to {strong-coupling} large Fr\"ohlich bipolarons).

In discussing the bipolaronic superconductivity as a candidate theory
for high-$T_C$ superconductivity one should mention the analysis \ch{by}
Mott in terms of {\em spin bipolarons}. Mott (1991) has emphasized how a
metal-insulator transition occurs in several of the high-temperature
superconductors, the metal being superconducting. In analyzing this
transition he suggests the formation of spin polarons and the
possibility that they combine to form bipolarons. Those bipolarons could
replace the Cooper pairs. Alexandrov {\it et al.} (1994) have
also analyzed  the non-degenerate gas of bosons above $T_C$.
In particular, the linear $T$-behaviour of the resistivity, above $T_C$,
was accounted for. Mott envisaged the
possibility of two different mechanisms for superconductivity depending
on the hole concentration (for one of the mechanisms, the bosons do not
overlap, for the other they do) and stressed the analogy of
high-T$_C$ superconductors to the behaviour of superfluid $^4$He.
The quantitative formulation of this fact goes back to
Alexandrov and Ranninger (1992) who used
a mapping of the electronic specific heat of high-T$_C$ superconductors
onto that of $^4$He to show that its $\lambda$-like temperature dependence
may be described in terms of charged bosons.

Spin-bipolaron model has been invoked by Alexandrov {\it et al.} (1994b)
to explain the experimental data
on the Hall effect and resistivity in underdoped YBa$_2$Cu$_3$O$_{7-\delta}                                                      $
(Carrington {\it et al.}, 1993) basing on the assumption that at zero
temperature a part of the spin bipolarons are in Anderson localized states
due to disorder. The agreement between theory and experiment exists at least
in the region of temperatures higher than the temperature $T^*$ at which
slope of resistivity changes. This is illustrated in Figs.\,14 and 15,
representing the temperature dependence of the Hall coefficient and of the
resistivity for different values of the degree of reduction in
YBa$_2$Cu$_3$O$_{7-\delta}$.

The Cooper pairing of small polarons has been regarded as a possible
mechanism of high-T$_C$ superconductivity ---
polaronic superconductivity, see e.~g. a  comprehensive review by
Alexandrov and Mott (1994) and an extensive list of references therein.
It was shown that there exists an intermediate region of the coupling
constant, where the energy of interaction between small polarons
(resulting from the interplay between the Coulomb repulsion and the
attraction due to their interaction with the lattice)
is smaller, or of the same order, as the small-polaron bandwidth,
both being less or comparable to the phonon frequency.
In this region, the BSC theory predicts superconductivity even for onsite
repulsion, provided that it is weaker compared to the
total intersite attraction of small polarons. It is the {\it polaronic
narrowing} of the band which increases the estimated maximum value for the
critical temperature when the interaction with high-frequency phonons is
dominant. This maximum occurs for the intermediate
value of the coupling constant which is situated at the boundary between the
above-described polaronic superconductivity region, on the one hand, and the
bipolaronic superconductivity region, on the other hand.
The latter mechanism is considered to be relevant for the values of the
coupling constant high enough to make the energy of attraction between
small polarons much larger than the small-polaron bandwidth.
In this case, the groundstate of the system of small polarons
can be --- in the first approximation --- regarded as consisting of a
phonon field and a set of immobile small {\it bipolarons};
the hopping term treated as a perturbation producing the
bipolaronic motion.

A serious alternative to the
picture developed by Alexandrov and Mott (1994) is the fermion-boson
model considering a possibility of exchange between localized bipolarons
and itinerant electrons which goes back to Ranninger and Robaszkiewicz
(1985). Depending on the filling, this model gives either a B'S-like
state or a superfluid state of bipolarons with a dramatic increase of
$T_C$ beyond a critical concentration.

In summary \ch{of this section}, it is  realized  that no definite theory for high-$T_C$
superconductivity  (whether or not it involves bipolarons)
is available as yet; nevertheless \ch{dielectric bipolarons},
spin-bipolaron gases and polaron-bipolaron mixtures seem to be systems of
significant promise in attempting to construct such a theory.

\section{Further Developments of the Polaron \\
 Concept}
\subsection{Polarons and bipolarons in polymers.
Solitons}

Many conjugated polymers (e.~g., trans-polyacetylene) behave as quasi-1D
semiconductors, where (\ch{the energy gaps} are between 1.5 and 3.0 eV).
The carrier transport in those
systems seems to take place through {\em charged defects} originating as
a result of doping. These charged defects are influenced by the
polaronic interaction with the surroundings. At low temperature the
transport is dominated by the motion of the defects; at higher doping
levels hopping conductivity occurs.
As is well known, the possible application of those polymers (including,
 e.~g., those in aerospace technology) have attracted wide attention.

Su {\it et al.} (1980) have proposed a model Hamiltonian to study
the physics of conjugated polymers. In analyzing the polaronic localized
states related to the charged defects non-linear behaviour,
characteristic of solitons, is \ch{revealed}.
The polaron-type excitations were experimentally
observed in polyacetylene by Su and Schreiffer (1980).

Polarons in conjugated polymers constitute a challenging and vast
subject; in the context of this article \ch{only} its
significance and main ingredients \ch{are pointed out};
the reader is referred to comprehensive reviews by Yu (1988) and
Fisher {\it et al.} (1989). Bipolarons might occur in polymers at
higher dopant levels and the question of polaron-bipolaron conversion
rates is a central one.

\subsection{\ch{Modelling systems using the polaron concept}}
Originally the polaron was \ch{mainly studied} to
describe the electron-LO pho\-non
interaction. The concept has been generalized to several systems where
one or many fermions interact with a bath of bosons or to instances
where such an approximation is meaningful.
\ch{Table 4} lists several such cases.
The fundamental concepts concerning a piezopolaron --- it is a name for a quasiparticle
arising due to piezoelectric interaction
between an electron and acoustic phonons ---
were displayed by Mahan (1972) and Lax (1972).
More recently, convincing experimental evidences on the significant
role of piezopolarons in the of the electronic transport properties
of CdS have been presented by Mahan (1990).

Other examples include the motion of a $^3$He atom through superfluid
helium (Bardeen {\it et al.}, 1969), or the dynamics of electrons (2DEG)
at the surface of liquid He --- ``ripplonic polaron''
(Jackson, Platzman, 1981; Devreese, Peeters, 1987).

Charges moving through liquids (many of which are characterized by
polarizable atoms) have been modelled as ``hydrated polarons''
(Laria {\it et al.}, 1991). Even an
electron interacting with plasmons (``plasmaron'') was invoked to study
the electron gas. A new subject concerns the occurrence of polarons in
fullerenes (Matus {\it et al.}, 1992).
\ch{Of special interest is the electronic polaron.}

These extensions of the polaron concept are not treated in any detail
here, but they illustrate the richness of the polaron concept.

A valuable extension of the polaron concept arises by considering the
interaction between a carrier and the exciton field. One of the early
formulations of this model was by Toyozawa (1963). The resulting
quasiparticle is called the {\em electronic polaron}.

The self-energy of the electronic polaron (which is almost independent
of wave number) must be taken into account when the bandgap of an
insulator or semiconductor is calculated using pseudopotentials. E. g. if
one calculates, with Hartree-Fock theory, the bandgap of an alkali halide,
one is typically off by a factor of two. This was the original problem
which was solved conceptually with the introduction of the electronic
polaron (Toyozawa, 1963; Kunz, 1974).
Also in the soft X-ray spectra of alkali halides
exciton sidebands have been observed which seems to be due to the
electronic polaron coupling (Devreese {\it et al.}, 1972).

The standard ``Local Density Approximation'' (LDA) does not provide
the correct bandgap for
semiconductors. The solution to this problem (Hybertsen, Louie, 1987;
Godby {\it et al.}, 1988) via the so-called
GW-approximation of Hedin (1965) is analogous in its basic interaction
to Toyozawa's solution of the gap problem for strongly ionic crystals.

\subsection*{Acknowledgement}
The author likes to express his gratitude to several colleagues for many
stimulating discussions and contributions on the polarons over the years:
R.~Evrard, F.~Peeters, F.~Brosens, L.~Lemmens. He also wishes to
acknow\-ledge recent interactions and fruitful discussions in particular
with V.~Fomin, and also with M.~Smondyrev.

This work has been partially supported by the E.~E.~C. Human Capital and
Mobility Program under contract No. CHRX-CT93-0124.

\newpage

\vskip 1truecm
{\Large \bf List of Works Cited}
\vskip 0.5truecm


  Adamowski, J. (1985),
{\underline {Phys. Rev.}} { B 32}, 2588 - 2595.

  Adamowski, J. (1989),
{\underline {Phys. Rev.}} { B 39}, 3649 - 3652.

  Adamowski, J., Gerlach, B., Leschke, H. (1981),
{\underline {Phys. Rev.}} { B 23}, 2943 - 2950.

  Aldrich, C., Bajaj, K. K. (1977),
{\underline {Solid State Commun.}} { 22}, 157 - 160.

  Alexandrov, A. S., Bratkovsky, A. M., Mott, N. F. (1994a),
{\underline {Phys. Rev.}}

\noindent {\underline {Lett.}} { 72}, 1734 - 1737.

  Alexandrov, A. S., Bratkovsky, A. M., Mott, N. F. (1994b),
{\underline {Physica}} {C 235---240}, 2345 - 2346.

  Alexandrov, A. S., Krebs, A. B. (1992),
{\underline {Usp. Fiz. Nauk}} {162}, 1 - 85

\noindent [English translation:
{\underline {Sov. Phys. Usp.}} { 35}, 345 - 383].

  Alexandrov, A. S.,  Mott N. F. (1994),
{\underline {Reports on Progress in Physics}} { 57}, 1197 - 1288.

  Alexandrov, A., Ranninger, J. (1981a)
{\underline {Phys. Rev.}} { B 23}, 1796 - 1801.

  Alexandrov, A., Ranninger, J. (1981b)
{\underline {Phys. Rev.}} { B 24}, 1164 - 1169.

  Alexandrov, A., Ranninger, J. (1982)
{\underline {Solid State Commun.}} { 81}, 403 - 406.

  Anderson, P. W. (1975),
{\underline {Phys. Rev. Lett.}} { 34}, 953 - 955.

  Appel, J.  (1968), in F. Seitz, D. Turnbull, H. Ehrenreich (Eds.),
{\underline {Solid Sta-}}

\noindent {\underline {te Physics}}, Vol. { 21}, 193 - 391.

  Austin, I. G., Mott, N. F. (1969),
{\underline {Advances in Physics}} {18}, 41 - 102.

  Bajaj, K. K. (1972), in
J. T. Devreese (Ed.),
{\underline {Polarons in Ionic Crystals}}
{\underline {and Polar Semiconductors}},
          Amsterdam: North-Holland,
p. 193 - 225.

  Bardeen, J., Baym, G., Pines, D. (1969),
{\underline {Phys. Rev.}} { 156}, 207 - 221.

  Bassani, F., Baldereschi, A. (1973),
{\underline {Surf. Sci.}} {37}, 304 - 327.

  Bassani, F., Pastori Parravicini, G. (1975),
{\underline {Electronic States and Optical}}

\noindent{\underline {Transitions in Solids}},
Oxford: Pergamon.

  Bassani, F., Geddo, M., Iadonisi, G., Ninno, D. (1991),
{\underline {Phys. Rev.}} { B 43}, 5296 - 5306.

  Benoit \'a la Guillaume, C. (1993),
{\underline {Phys. Stat. Sol.}} {(b) 175}, 369 - 380.

  Bogolubov, N. N. (1950),
{\underline {Ukr. Matem. Zh.}} { 2}, 3.

  Bogolubov, N. N., Tyablikov, S. V. (1949),
{\underline {Zh. Eksp. i Teor. Fiz.}} { 19}, 256.

  Brandt, R. C., Brown, F. C. (1969),
{\underline {Phys. Rev.}} { 181}, 1241 - 1250.

  Brosens, F., Devreese, J.~T. (1988),
{\underline {Phys.~Stat.~Sol.}} {(b) 145}, 517 - 523.

  Brosens, F., Lemmens, L., Devreese, J. T. (1991),
{\underline {Phys. Rev.}} { B 44}, 10296 - 10299.

  Brown, F. C. (1963), in
C. G. Kuper, G. D. Whitfield (Eds.) (1963),
{\underline {Polarons and Excitons}}, Edinburgh: Oliver and Boyd,
p. 323 - 355.

 Brown, F. C. (1981), in
J. T. Devreese (Ed.) (1981), {\underline {Recent Develop-}}
{\underline {ments in Condensed Matter Physics}}, Vol. 1, New York:
Plenum, p. 575 - 591.

  Carrington, A., Walker, D. J. C., Mackenzie, A. P.,
Cooper, J. R. (1993),
{\underline {Phys. Rev.}} {B 48}, 13051 - 13059.

  Casado, J. M., Harding, J. H., Hyland, G. J. (1994),
{\underline {J. Phys.: Condens.}}

\noindent{\underline {Matter}} { 6}, 4685 - 4698.

  Cataudella, V., Iadonisi, G., Ninno, D. (1992),
{\underline {Europhysics Letters}} { 17}, 709 - 714.

%

  Chakraverty, B. K., Feinberg, D., Hang, Z., Avignon, M.
(1987), {\underline {Solid}}

\noindent{\underline {State Commun.}} { 64}, 1147 - .

  Cheng, J.-P., McCombe, B. D., Brozak, G., Schaff, W. (1993),
        {\underline {Phys.~Rev.}} {B 48}, 17243 - 17254.

  Cohen, M. H., Economou, E. N., Soukoulis C. M. (1984),
{\underline {Phys. Rev.}} {B 29}, 4496 - 4499; 4500 - 4504.

  Cohn, D. R., Larsen, D. M., Lax, B. (1972),
{\underline {Phys.~Rev.}} {B 6}, 1367 - 1375.

  Devreese, J. T., (1963),
{\underline {Bull.\ Belgian Physical Society}} III, 259 - 263.

  Devreese, J. T. (Ed.) (1972),
{\underline {Polarons in Ionic Crystals and Polar Semi-}}

\noindent {\underline {conductors}},
Amsterdam: North-Holland.

%
  Devreese, J.~T.,  Brosens, F. (1992),
{\underline {Phys.~Rev.}} {B 45}, 6459 - 6478.

Devreese, J. T., Evrard, R. (1966), {\underline {Phys. Letters}}
{ 11}, 278 - 279.

  Devreese, J., Evrard, R., Kartheuser, E., Brosens, F. (1982),
{\underline {Sol. Stat.}}

\noindent {\underline {Commun.}} { 44}, 1435 - 1438.

  Devreese, J. T., Kunz, A. B., Collins, T. C. (1972),
{\underline {Solid State Commun.}} { 11}, 673 - 678.

  Devreese, J. T., Peeters, F. (Eds.) (1984),
{\underline {Polarons and Excitons in Po-}}

\noindent{\underline {lar Semiconductors and Ionic Crystals}},
New York: Plenum.

  Devreese, J. T.,  Peeters, F. M. (Eds.) (1987),
{\underline {The Physics of the Two-Di-}}
{\underline {mensional Electron Gas}},
New York: Plenum.


  Emin, D., Hillery, M. S. (1989),
{\underline {Phys. Rev.}} { B 39}, 6575 - 6593.

  Emin, D., Holstein, T., (1976),
{\underline {Phys. Rev. Lett.}} { 36}, 323 - 326.

  Engineer, M. H., Tzoar, N. (1972), in
J. T. Devreese (Ed.),
{\underline {Polarons in}}

\noindent{\underline { Ionic Crystals and Polar
Semiconductors}},
          Amsterdam: North-Holland,
p. 747 - 754.

  Evrard, R. (1965),
{\underline {Phys. Letters}} { 14}, 295 - 296.

  Evrard, R., Kartheuser, E., Devreese, J.~T. (1970),
        {\underline {Phys.~Stat.~Sol.}} {(b) 41}, 431 - 438.

  Falck, J. P., Levy, A., Kastner, M. A., Birgeneau, R. J. (1992),
{\underline {Phys. Rev.}}

\noindent {\underline {Lett.}} { 69}, 1109 - 1112.

  Feynman, R. P. (1955), {\underline {Phys. Rev.}} { 97}, 660 - 665.

  Feynman, R. P. (March 1973), private communication.

  Feynman, R. P., Hibbs, A. R. (1965), {\underline {Quantum Mechanics and
Path In-}}

\noindent {\underline {tegrals}}, McGraw-Hill: New York, p. 308.

  Feynman, R. P., Hellwarth, R. W., Iddings, C. K.,
Platzman, P. M. (1962),
{\underline {Phys. Rev.}} { 127}, 1004 - 1017.

  Finkenrath, H., Uhle, N., Waidelich, W. (1969),
{\underline {Solid State Commun.}} { 7}, 11 - 14.

%

  Firsov, Yu. A. (Ed.) (1975) {\underline {Polarons}},
Moscow: Nauka.

  Fisher, A. O., Hayes, W., Wallace, D. S. (1989),
{\underline {J. Phys.: Condens. Mat-}}

\noindent {\underline {ter}} { 1}, 5567 - 5593.

  Fomin, V. M., Pokatilov, E. P. (1988),
{\underline {Physics Reports}}, 158, 205 - 336.

  Fomin, V. M., Smondyrev, M. A. (1994),
{\underline {Phys. Rewv.}} {B 49}, 12748 - 12753.

  Friedman, L., Holstein, T. (1963),
{\underline {Ann. Phys.}} { 21}, 474 - 549.

  Fr\"ohlich, H. (1937),
{\underline {Proc. Roy. Soc.}} {A 160}, 230 - .

  Fr\"{o}hlich, H. (1954),
{\underline {Advances in Physics}} { 3}, 325 - .

  Fr\"ohlich, H. (1957),
{\underline{Arch.\ Sci.\ Gen\`eve}} {10}, 5 - 6.

  Fr\"ohlich, H., Sewell, G. L. (1959),
{\underline{Proc.\ Phys.\ Soc.}} {74}, 643 - 647.

  Gehlig, R., Salje, E. (1983),
{\underline {Phil. Mag}} 47, 229 - 245.

  de Gennes, P. G. (1960),
{\underline {Phys. Rev.}} { 118}, 141 - 154.

  Godby, R. W., Schl\"uter, M., Sham, L. J. (1988),
{\underline {Phys. Rev.}} { B 37}, 10159 - 10175.

  Gurevich, V. L., Lang, I. G., Firsov, Yu. A. (1962),
{\underline {Fiz. Tverd. Tela}} { 4}, 1252 -
[English translation: (1963),
{\underline {Sov. Phys.---Solid State}} { 4}, 918 - ].

  Haken, H. (1956),
{\underline {Il Nuovo Cimento}} { 3}, 1230 - .

 Hashimoto, T., Hirasawa, R., Yoshida, T., Yonemura, Y., Mizusaki, J.,
 Tagawa, H. (1995),
{\underline {Phys. Rev.}} { B 51}, 576 - 580.

  Hedin, L. (1965),
{\underline {Phys. Rev.}} { 139}, A796 - A823.

  Hellwarth, R.~W., Platzman, P.~M. (1962),
{\underline {Phys.~Rev.}} {128}, 1599 - 1604.

  Hiramoto, H., Toyozawa, Y.  (1985),
{\underline {J. Phys. Soc. Japan}} { 54}, 245 - 259.

  Hodby, J.~W., Russell, G., Peeters, F., Devreese,
        J.~T., Larsen, D.~M. (1987),
{\underline {Phys.~Rev.~Lett.}} { 58}, 1471 - 1474.

 H\"ohler, G., M\"ullensiefen, A. (1959),
{\underline {Z. Phys.}}, {157}, 159 - 165.

  Holstein, T. (1959),
{\underline {Ann. Phys.}} (USA), 8, 343 - 389.

  Howarth, D. J.,  Sondheimer, E. H. (1953),
{\underline {Proc. Roy. Soc.}} { A 219}, 53 - .

  Hubbard, J. (1964),
{\underline {Proc. Roy. Soc.}} {A 281}, 401 - .

  Hybertsen, M. S., Louie S. G. (1987),
{\underline {Phys. Rev.}} { B 35}, 5585 - 5601;
5602 - 5610.

  Iadonisi, G., Cataudella, V., Ninno, D., Chiofalo, M. L. (1995),
{\underline {Phys.}}

\noindent {\underline {Letters}} {A 196}, 359 - 364.

  Jackson, S. A., Platzman, P. M. (1981),
{\underline {Phys. Rev.}} {B 24}, 499 - 502.

  Johnson, E., Larsen, D. (1966),
{\underline {Phys.~Rev.~Lett.}} { 16}, 655 - 659.

 de Jongh, L. J. (1988),
{\underline {Physica C}} { 152}, 171 - 216.

 Jung, W. H., Iguchi, E. (1995),
{\underline {J. Phys.: Condens. Matter}} {7}, 1215 - 1227.

  Kadanoff, L. P. (1963),
{\underline {Phys. Rev.}} { 130}, 1364 - 1369.

  Kartheuser, E., Devreese, J., Evrard, R. (1979),
{\underline {Phys. Rev.}} {B 19}, 546 - 551.

  Kartheuser, E., Evrard, R., Devreese, J. (1969),
{\underline {Phys. Rev. Lett.}} { 22}, 94 - 97.

  Klinger, M. I. (1979),
  {\underline {Problems of Linear Electron (Polaron) Trans-}}

\noindent {\underline  {port Theory in Semiconductors}},
Oxford: Pergamon Press.

  Knox, R. (1963), {\underline {Theory of Excitons}},
New York: Academic Press.

  Kunz, A. B. (1974), in
J. T. Devreese, A. B. Kunz, T. C. Collins (Eds.)
{\underline {Elementary Excitations in Solids, Molecules
and Atoms}},
Vol. A, New York: Plenum, p. 159 - 187.

  Kuper, C. G., Whitfield, G. D. (Eds.) (1963),
{\underline {Polarons and Excitons,}} Edinburgh: Oliver and Boyd.

  Landau, L. D. (1933),
{\underline {Phys. Z. Sovjet.}} { 3}, 664
[English translation:

\noindent (1965), {\underline {Collected Papers}},
New York: Gordon and Breach, p. 67 - 68].

  Lang, I.,  Firsov, Yu. (1962),
{\underline {Zh. Eksp. i Teor. Fiz.}} { 43},
        1843 - 1860
[English translation:
(1963),
{\underline {Sov. Phys.---JETP}} { 16}, 1301 - 1312].

  Lannoo, M., Baraff, G. A., Schl\"uter, M. (1959),
{\underline {Phys. Rev.}} {B 24}, 955 - 963.

  Laria, D., Wu, D., Chandler, D. (1991),
{\underline {J. Chem. Phys.}} { 95}, 4444 - 4453.

  Larsen, D. M. (1969),
{\underline {Phys. Rev.}} { 187}, 1147 - 1152.

  Larsen, D. (1972), in
J. T. Devreese (Ed.),
{\underline {Polarons in Ionic Crystals and}}
 {\underline {Polar Semiconductors}},
          Amsterdam: North-Holland,
p. 237 - 287.

  Larsen, D. M. (1981),
{\underline {Phys. Rev. B}} { 23}, 628 - 631.

  Larsen, D. (1984a),
{\underline {Phys.~Rev.}} { B 30}, 4595 - 4608.

  Larsen, D. (1984b),
{\underline {Phys.~Rev.}} { B 30}, 4807 - 4808.

  Larsen, D.~M. (1985),
{\underline { Phys.~Rev.}} { B32}, 2657 - 2658.

  Larsen, D. (1991), in: G. Landwehr and E. Rashba (Eds.),
{\underline {Landau Level}}
 {\underline {Spectroscopy}}, Vol. 1, Amsterdam: North Holland,
p. 109 - 130.

  Lax, B. (1972), in
J. T. Devreese (Ed.),
{\underline {Polarons in Ionic Crystals and}}
 {\underline {Polar Semiconductors}},
          Amsterdam: North-Holland,
p. 755 - 782.

  Lee, T. D., Low, F. E., Pines, D.  (1953),
{\underline {Phys. Rev.}} { 90}, 297 - 302.

  L\'{e}pine, Y., Matz, D. (1976),
{\underline {Can.~J.~Phys.}} { 54}, 1979 - 1989.

  Lerner, R. G., Trigg, G. L. (Eds.) (1991)
{\underline {Encyclopedia of Physics}}, New York: VCH, p. 941.

  Lieb, E.  (1977),
{\underline {Studies in Applied Mathematics}} { 57}, 93 - .

  Litton, C. W., Button, K. J., Waldman, J., Cohn, D. R.,
Lax, B. (1976),
{\underline {Phys.~Rev.}} {B 13}, 5392 - 5396.

  Mahan, G. D. (1972), in
J. T. Devreese (Ed.),
{\underline {Polarons in Ionic Crystals}}
 {\underline {and Polar Semiconductors}},
          Amsterdam: North-Holland,
p. 553 - 657.

  Mahan, G. D. (1990),
{\underline  {Many-Particle Physics}}, New York: Plenum, pp. 39 - 40;
635 - 636.

  Marshall, J.~T., Chawla, M. (1970),
{\underline {Phys.~Rev.}} {B 2}, 4283 - 4287.

  Masumi, T., Minami, H.,  Shimada, H. (1988a),
{\underline {J. Phys. Soc. Japan}} { 57}, 2674 - 2677.

  Masumi, T., Shimada, H., Minami, H. (1988b),
{\underline {J. Phys. Soc. Japan}} { 57}, 2670 - 2673.

  Matus, M., Kuzmany, H., Sohmen, E. (1992),
{\underline {Phys. Rev. Lett.}} {68}, 2822 - 2825.

  Merkt, U. (1985),
{\underline {Phys.~Rev.}} {B 32}, 6699 - 6712.

  Micnas, R., Ranninger, J., Robaszkiewicz, S. (1990),
{\underline {Rev. Mod. Phys.}} { 62}, 113 - 171.

  Miura, N., Nojiri, H., Imanaka, Y. (1994), in D. J. Lockwood (Ed.),
{\underline {22nd International
Conference on the
Physics of Semiconductors}}, Vol. 2,

\noindent Singapore: World Scientific, 1111 - 1118.

  Miyake, S. J. (1975),
{\underline {J. Phys. Soc. Japan}} { 38}, 181 - 182.

  Mott, N. F. (1987),
{\underline {Conduction in Non-Crystalline Materials}},
Oxford: Clarendon.

  Mott, N. F. (1990),
{\underline {Metal-Insulator Transitions}}, London: Taylor and Francis.

  Mott, N. F. (1991), in
D. P. Tunstall, W. Barford (Eds.),
{\underline {High Tempera-}}

\noindent{\underline {ture Superconductivity}},
         Bristol: Adam Hilger, p. 271 - 294.

  Mott, N. F.,  Davis, E. A. (1979),
{\underline {Electronic Processes in  Non-Crystalline }}

\noindent{\underline {Materials}},
Oxford: Clarendon.

 Nagels, P., Devreese, J., Denayer, M. (1964),
{\underline {J. Appl. Phys.}} {35}, 1175 - .

  Nicholas, R. J., Watts, M., Howell, D. F., Peeters, F. M.,
Xiaoguang, Wu, Devreese, J. T., van Bockstal, L., Herlach, F.,
Langerak, C. J. G. M., Singleton, J., Chery, A. (1992),
{\underline {Phys. Rev.}} { B 45}, 12144 - 12147.

%
%
  Peeters, F.~M., Devreese, J.~T. (1982),
{\underline {Phys.~Rev.}} {B 25}, 7281 - 7301.

  Peeters, F., Devreese, J. T. (1984), in:
F. Seitz, D. Turnbull (Eds.),
{\underline {Solid State Physics}}, {Vol. 38}, p. 81 - 133.

%
  Peeters, F.~M., Devreese, J.~T. (1985),
{\underline {Phys.~Rev.}} { B 31}, 3689 - 3695.

  Peeters, F.~M., Devreese, J. T. (1986),
{\underline {Phys.~Rev.}} {B 34}, 7246 - 7259.

  Peeters, F.~M. and Devreese, J.~T. (1987),
{\underline {Phys.~Rev.}} { B 36}, 4442 - 4445.

  Peeters, F.~M. Xiaoguang, Wu,  Devreese, J.~T. (1986a),
        {\underline {Phys.~Rev.}} {B 33}, 3926 - 3934.

  Peeters, F.~M., Xiaoguang, Wu,  Devreese, J.~T. (1986b),
{\underline {Phys.~Rev.}} { B 33}, 4338 - 4340.

  Peeters, F.~M., Xiaoguang, Wu,  Devreese, J.~T. (1986c),
{\underline {Phys.~Rev.}} { B 34}, 1160 - 1164.

  Peeters, F.~M., Xiaoguang, Wu, Devreese, J.~T. (1988a),
{\underline {Surf.~Sci.}} { 196}, 437 - .

  Peeters, F.~M., Xiaoguang, Wu, Devreese, J.~T. (1988b),
{\underline{Solid State}}

\noindent {\underline {Commun.}} {65}, 1505 - 1508.

%
  Pekar, S. I. (1951), {\underline {Research in
Electron Theory of Crystals}}, Moscow: Gostekhizdat
[German translation: (1954),
{\underline {Untersuchungen}}

\noindent {\underline  {\"{u}ber die
Elektronentheorie der Kristalle}}, Berlin: Akademie Verlag;
English translation: (1963), {\underline {Research in
Electron Theory of Crystals}}, US AEC Report AEC-tr-5575].

  Petelenz, P., Smith, Jr., V. H. (1981),
{\underline {Phys. Rev. B}} { 23}, 3066 - 3070.

  Platzman, P. M. (1962),
{\underline {Phys. Rev.}} { 125}, 1961 - 1965.

  Pollman, J., B\"uttner, H. (1975),
{\underline {Solid State Commun.}} { 17}, 1171 - 1174.

  Pollman, J., B\"uttner, H. (1977),
{\underline {Phys. Rev. B}} { 16}, 4480 - 4490.

  Ranninger, J. (1994),
{\underline {Physica}} {C 235 --- 240}, 277 - 280.

  Ranninger, J., Robaszkiewicz, S. (1985),
{\underline {Physica}} {B 135}, 468 - .

  Reik, H. G. (1972), in
J. T. Devreese (Ed.),
{\underline {Polarons in Ionic Crystals}}

\noindent{\underline {and Polar Semiconductors}},
          Amsterdam: North-Holland,
p. 679 - 714.

  Reik, H. G.,  Heese, D. (1967),
{\underline {J. Phys. Chem. Solids}} { 28}, 581 - 596.

 R\"oseler, J. (1968),
{\underline {Phys. Stat. Sol.}} {25}, 311 - 316.

  R\"uscher, C., Salje, E., Hussain, A. (1988),
{\underline {J. Phys. C: Solid State}}

\noindent {\underline {Physics}} 21, 3737 - 3749.

  Sak, J. (1972),
{\underline {Phys.~Rev.}} {B 6}, 3981 - 3986.

  Salje, E., G\"uttler B. (1984),
{\underline {Phil. Mag.}} B 50, 607 - 620.

  Das Sarma, S. (1984),
{\underline {Phys.~Rev.~Lett.}} { 52}, 859 - 862.

  Das Sarma, S., Mason, B. A. (1985),
{\underline {Ann. Phys.}}  (USA) {163}, 78 - .

  Schafroth, M. R. (1955),
{\underline {Phys. Rev.}} { 100}, 463 - 475.

  Scholz, J., Koch, F., Ziegler, J., Maier, H. (1983),
{\underline {Solid State Commun.}} { 46}, 665 - 668.

  Selyugin, O. V., Smondyrev, M. A. (1989),
{\underline {Phys. Stat. Sol.}} (b) {155}, 155 - 167.

  Seidenbush, W., Lindemann, G., Lassnig, R., Edlinger,
        J., Gornik, E. (1984),
{\underline {Surf.~Sci.}} { 142}, 375 - .

  Sewell, G. L. (1958),
{\underline{Phil.\ Mag.}} {3}, 1361 - 1380.

%
    Shi, J. M., Peeters, F. M., Devreese, J. T. (1995), in
{\underline {March Meeteing of}}

\noindent {\underline {the APS}}, San Jos\'e 1995 (to appear
in the Bulletin of the APS).

  Sigg, P., Wyder, P., Perenboom, J.~A.~A.~J. (1985),
        {\underline {Phys.~Rev.}} { B 31}, 5253 - 5261.

  Smondyrev, M. A. (1986),
{\underline {Teor. Math. Fiz.}} {68}, 29 - 44

\noindent
[English translation: {\underline {Theor. Math. Phys.}} {68}, 653].

  Smondyrev, M. A., Devreese, J. T., Peeters, F. M. (1995),
{\underline {Phys. Rev.}} {B 51}, 8 pages.

  Spicci,\,M., Salkola,\,M.\,I., Bishop,\,A.\,R. (1994),
{\underline {J. Phys.: Condens. Matter}} { 6}, L361 - L366.

  Stoneham, A. M. (1979)
{\underline {Advances in Physics}} { 28}, 457 - .

  Stoneham, A. M.,  Bullough, R. (1971),
{\underline {J. Phys.}} {C 3}, L195 - L197.

  Su, W. P., Schreiffer, J. R. (1980),
{\underline {Proc. Natl. Acad. Sci. USA}} {77}, 5626 - 5629.

  Su, W. P., Schrieffer, J. R., Heeger, A. J. (1980),
{\underline {Phys. Rev.}} {B 22}, 2099 - 2111.

  Tyablikov, S. V. (1951),
{\underline {Zh. Eksp. i Theor. Phys.}} {21}, 377.

%
%
%
  Toyozawa, Y. (1963),
in C. G. Kuper, G. D. Whitfield (Eds.) (1963),
{\underline {Polarons and Excitons}}, Edinburgh: Oliver and Boyd,
p. 211 - 232.

  Toyozawa, Y. (1964),
{\underline {J. Phys. Chem. Solids}} {25}, 59 - .

  Van Royen, J., Devreese, J.~T. (1981),
{\underline {Solid State Commun.}} { 40}, 947 - 949.

  Verbist, G., Peeters, F. M., Devreese, J. T. (1990),
{\underline {Sol. State. Commun.}} { 76}, 1005 - 1007.

  Verbist, G., Peeters, F. M., Devreese, J. T. (1991),
{\underline {Phys. Rev.}} { B 43}, 2712 - 2720.

  Verbist, G., Peeters, F. M., Devreese, J. T. (1994), in
Proceedings

\noindent{\underline {Workshop
on Polarons and Bipolarons in High-T$_C$ Superconductors
and}}

\noindent{\underline {Related Materials}}, Cambridge 1994 (in press).

  Vigneron, J., Evrard, R.~, Kartheuser, E. (1978),
        {\underline {Phys.~Rev.}} {B 18}, 6930 - 6943.

  Vinetskii, V. L. (1961),
{\underline {Zh. Eksp. i. Teor. Fiz.}} { 40}, 1459 - 1468

\noindent [English translation: {\underline {Sov. Phys.---JETP}} {13},
1023 - 1028].

  Waldman, J., Larsen, D. M., Tannenwald, P. E., Bradley, C. C.,
Cohn, D. R., Lax, B., (1969),
{\underline {Phys.~Rev.~Lett.}} { 23}, 1033 - 1037.

  Wallis, R. F., Balkanski, M. (1986),
{\underline {Many-Body Aspects of Solid State}}

\noindent {\underline {Spectroscopy}},
Amsterdam: North-Holland.

  Wood, R. F. (1991),
{\underline {Phys. Rev. Lett.}} { 66}, 829 - 832.

Xiaoguang, Wu, Peeters, F. M., Devreese, J. T. (1985),
        {\underline {Phys.~Rev.}} {B 31}, 3420 - 3426.

  Xiaoguang, Wu, Peeters, F.~M., Devreese, J.~T. (1987),
{\underline {Phys.~Rev.}} { B 36}, 9760 - 9764.

  Yamashita, J.,  Kurosawa, T. (1958),
{\underline {J. Phys. Chem. Solids}} { 5}, 34 - 43.

  Yamazaki, K. (1983),
{\underline {J. Phys.}} {A 16}, 3675 - 3685.

  Yu, Lu (1988),
{\underline {Solitons \& Polarons in Conducting Polimers}},
Singapore:
World Scientific, Chapter IV. Polarons and Bipolarons;
Chapter V. Soliton and Polaron Dynamics.


\newpage

{\Large \bf Further Reading}
\vskip 1truecm

\indent  Alexandrov, A. S.,  Mott N. F. (1994),

 {\underline {Reports on Progress in Physics}} { 57}, 1197 - 1288.
\medskip

  Appel, J.  (1968), in F. Seitz, D. Turnbull, H. Ehrenreich (Eds.),

 {\underline {Solid State Physics}}, Vol. { 21}, 193 - 391.
\medskip

 Bogolubov, N. N. (1970),

{\underline {Collected Papers}}, Vol. { 2}, Kiev: Naukova Dumka.
\medskip

  Devreese, J. T. (Ed.) (1972),

 {\underline {Polarons in Ionic Crystals and Polar
Semiconductors}},

  Amsterdam: North-Hol\-land.
\medskip

  Devreese, J. T., Peeters, F. (Eds.) (1984),

  {\underline {Polarons and
Excitons in Polar Semiconductors and Ionic Crystals}},

  New York: Plenum.
\medskip

  Devreese, J. T., Peeters, F. M. (Eds.) (1987),

  {\underline {The Physics
of the Two-Dimensional Electron Gas}}, New York: Plenum,

Chapter II. Electron-Phonon Interaction; Chapter IV. Special Topics.
\medskip

  Feynman, R. P., (1972),

  {\underline {Statistiucal Mechanics}},
Benjamin: Massachusetts.
\medskip

  Feynman, R. P., Hibbs, A. R. (1965),

  {\underline {Quantum Mechanics and Path Integrals}},
McGraw-Hill: New York.
\medskip

  Firsov, Yu. A. (Ed.) (1975),

  {\underline {Polarons}}, Moscow: Nauka.
\medskip

  Fisher, A. O., Hayes, W., Wallace, D. S. (1989),

 {\underline {J. Phys.: Condens. Matter}} { 1}, 5567 - 5593.
\medskip

  Fomin, V. M., Pokatilov, E. P. (1988),

  {\underline {Physics Reports}}, 158, 205 - 336.
\medskip

  Klinger, M. I. (1979),

  {\underline {Problems of Linear Electron (Polaron) Transport
Theory in}}

 {\underline  {Semiconductors}},  Oxford: Pergamon Press.
\medskip

  Kuper, C. G., Whitfield, G. D. (Eds.) (1963),

  {\underline {Polarons and Excitons}}, Edinburgh: Oliver and Boyd.
\medskip

  Lakhno, V. D. (Ed.) (1994),

 {\underline {Polarons \& Applications (Proceedings in nonlinear
science)}}, Chi\-ches\-ter:

John Wiley \& Sons.
\medskip

  Yu, Lu (1988),

  {\underline {Solitons \& Polarons in Conducting Polimers}},

  Singapore: World Scientific, Chapter IV. Polarons and Bipolarons;

Chapter V. Soliton and Polaron Dynamics.
\medskip

  Mott, N. F. (1990),

{\underline {Metal-Insulator Transitions}}, London: Taylor and Francis.
\medskip

  Pekar, S. I. (1951),

{\underline {Research in Electron Theory of Crystals}}, Moscow: Gostekhizdat

[German translation: (1954), {\underline {Untersuchungen
\"{u}ber die
Elektronentheorie}}

{\underline {der Kristalle}}, Berlin: Akademie Verlag;
English translation: (1963),

{\underline {Research in
Electron Theory of Crystals}}, US AEC Report AEC-tr-5575].


\newpage
                        \begin{table}[t]
                        \caption{{\em Fr\"{o}hlich coupling
                        constants}}
			   \bigskip
\begin{center}
                        \begin{tabular}{llrr}
                        \hline
                        \multicolumn{1}{c}{Material} &
                        \multicolumn{1}{c}{$\alpha$} &
                        \multicolumn{1}{c}{Material} &
                        \multicolumn{1}{c}{$\alpha$}
                       \\ \hline
                         CdTe  &  0.31   & KI     & 2.5          \\
                         CdS   &  0.52   & RbCl   & 3.81         \\
                         ZnSe  &  0.43   & RbI    & 3.16         \\
                         AgBr  &  1.6    & CsI    & 3.67         \\
                         AgCl  &  1.8    & TlBr   & 2.55         \\
                         CdF$_{2}$& 3.2    & GaAS   & 0.068         \\
                         InSb  &  0.02   & GaP    & 0.201         \\
                         KCl   &   3.5   & InAs   & 0.052         \\
                         KBr   &   3.05  & SrTiO$_{3}$ & 4.5       \\ \hline
                        \end{tabular}
\end{center}
\vskip 10cm
                        \end{table}

\newpage
                        \begin{table}[t]
                        \caption{\em Groundstate energy (in units
                           $\hbar \omega_{\rm LO}$) of the bound polaron
                         for several values of $\alpha$ and ${\tilde \beta}$ as
                          obtained by Devreese {\it et al.} (1982) compared to
                           the variational results by Larsen (1969)}
\bigskip
\begin{center}
                        \begin{tabular}{r|ll|ll}
                        \hline
                          \multicolumn{1}{r|}{}
                        & \multicolumn{2}{c|}{} &
                         \multicolumn{2}{c}{}  \\
                          \multicolumn{1}{r|}{$\alpha$}
                        & \multicolumn{2}{c|}{${\tilde \beta}=6.32$} &
                         \multicolumn{2}{c}{${\tilde \beta}=4.47$}  \\
			     \multicolumn{1}{r|}{}
                        & \multicolumn{1}{c}{E\ (Devreese {\it et al.},} &
                          \multicolumn{1}{c|}{E\ (Larsen,} &
                          \multicolumn{1}{c}{E\ (Devreese {\it et al.},} &
                          \multicolumn{1}{c}{E\ (Larsen,}\\
                        & \multicolumn{1}{c}{1982)} &
                          \multicolumn{1}{c|}{1969)} &
                          \multicolumn{1}{c}{1982)} &
                          \multicolumn{1}{c}{1969)}
                       \\ \hline
                          2 & $-$14.66 &  $-$14.69 & $-$8.60 & $-$8.64 \\
                          5 & $-$23.0  &  $-$23.0  & $-$15.21 & $-$15.30 \\
                          7 & $-$29.41 &  $-$29.47 & $-$20.52 & $-$20.62 \\
                          11 & $-$44.6 & $-$44.6 & $-$33.4 & $-$33.4 \\ \hline
                        \end{tabular}
\end{center}
\vskip 10cm
                        \end{table}


\newpage
                        \begin{table}[t]
                        \caption{\em Calculated energy of the first
                          excited 2p state (in units
                           $\hbar \omega_{\rm LO}$) of the bound polaron
                         for several values of $\alpha$ and ${\tilde \beta}$ as
                          obtained by Devreese {\it et al.} (1982)}
\bigskip
\begin{center}
                        \begin{tabular}{rlll}
                        \hline
                         \multicolumn{4}{c}{E$_{\rm 2p}$}  \\
                         \hline
                         {} &\ \ {}\ \  &\ \  {} \ \ & \ \ {} \\
                         $\alpha$ &\ \ ${\tilde \beta}=0$\ \  &\ \  ${\tilde \beta}=1$ \ \ &
                                     \ \ ${\tilde \beta}=2$ \\
                          \hline
                          1 & $-$0.7626 & $-$0.8775 & $-$1.113 \\
                          3 & $-$1.971  & $-$2.237  & $-$2.628 \\
                          5 & $-$3.207  & $-$3.644  & $-$4.205 \\
                          7 & $-$4.600  & $-$5.214  & $-$5.955  \\
                          9 & $-$6.199  & $-$6.996  & $-$7.922  \\
                          11 & $-$8.029 & $-$9.014  & $-$10.13  \\ \hline
                        \end{tabular}
\end{center}
\vskip 10cm
                        \end{table}

\newpage
                        \begin{table}[t]
                        \caption{{\em Extensions of the
                        polaron concept}}
			   \bigskip
\begin{center}
                        \begin{tabular}{llrr}
                        \hline
                        \multicolumn{1}{c}{Concept} &
                        \multicolumn{1}{c}{``Candidates''} &
                       \\ \hline
                       Acoustic polaron      & Hole in AgCl   \\
                       Piezoelectric polaron & ZnO, ZnS, CdS\\
                       Electronic polaron    & Bandstructure of all \\
                                             &  semiconductors\\
                       Spin polaron          & Magnetic semiconductors\\
                       Bipolarons:            & \\
                       \hspace*{0.3truecm} One center & Amorphous
                       chalcogenides\\
                       \hspace*{0.3truecm} Two center & Ti$_{4}$O$_{7}$,
                       vanadium-bronzes \\
                        Small polarons
                       and hopping &
                        Transition metal oxydes \\
                                 & Polyacetylenes \\
                        Superconductivity &
                         BaPb$_{1-x}$ (1981)   \\
                        and (bi)-polarons
                          &  Bi$_{x}$O$_{3}$   \\
                         (``Localized
                        electron pairs'') &
                          LiTi$_{2}$O$_{4}$ (1985) \\
                          & ``High-T$_{C}$ superconductivity'' \\ \hline
                        \end{tabular}
\end{center}
\vskip 10cm
                        \end{table}


\newpage
\begin{center}
Figure captions
\end{center}

{Fig.\,1. A conduction electron (or hole) together with its
self-induced polarisation in a polar semiconductor or an
ionic crystal forms a quasiparticle: a polaron.}

{Fig.\,2. Internal excitations at strong coupling: $E_0$ --- groundstate,
$E_1$ --- first relaxed excited state; $E_{FC}$ --- Franck-Condon state.}

{Fig.\,3. Polaron mobility in AgBr (solid line) according to the low
temperature self consistent approach of
Kartheuser, Devreese and Evrard (1979)
compared with the Hall data, from (Brown, 1981).}

{Fig.\,4. Optical absorption $\protect\Gamma$ of polarons at $\alpha=6$
as a function of frequency $\nu$, expressed in units $\omega_{\rm LO}$,
from (Devreese, 1972).}

{Fig.\,5. The cyclotron resonance position plotted as a function of
magnetic field for InSe from (Nicholas {\it et al.}, 1992).}

{Fig.\,6. Induced absorption in AgBr at 9.3K from 150 to 320 cm$^{-1}$.
The exciting light was in the region of the direct absorption edge.
From (Brandt, Brown, 1969);
reproduced by courtesy of the American Physical Society.}

{Fig.\,7. Real part of the conductivity versus normalized frequency
$\nu\tau$ for different values of the parameter $\Gamma$
from (Reik, Heese, 1967);
reproduced by courtesy of the Pergamon Press Ltd.}


{Fig.\,8. Integral intensities of the polaronic absorption in dependence on
the degree of reduction of NbO$_{2.5-x}$ ($\bullet$)
and WO$_{3-x}$ ($\Box$) from (R\"uscher {\it et al.}, 1988);
reproduced by courtesy of the Insitute of Physics Publishing.}

{Fig.\,9. Seebeck coefficient of BaBi$_{0.25}$Pb$_{0.75}$O$_{3.00}$:
measured data ($\bullet$) and phenomenoligical estimates (solid and
dashed lines). From (Hashimoto {\it et al.}, 1995);
reproduced by courtesy of the American Institute of Physics.}

{Fig.\,10.
Arrhenius plot for conductivity in
La$_{1-x}$Sr$_{x}$FeO$_3$ with
$x = 0.05$ ($\bullet$),
$x = 0.10$ ($\circ$),
$x = 0.20$ (solid boxes),
$x = 0.25$ ($\Box$),
$x = 0.30$ ($\diamond$) in comparison with small-polaron theory
(solid lines). From (Jung, Iguchi, 1995);
reproduced by courtesy of the Insitute of Physics Publishing.}

{Fig.\,11. The stability region for bipolaron formation in 3D
from (Verbist {\it et al.}, 1990).
The dotted line $U=\protect\sqrt{2}\alpha$ separates the physical
region
$(U\geq\protect\sqrt{2}
\alpha)$ from the non-physical
$(U\leq\protect\sqrt{2}\alpha)$. The stability region lies below
the full curve. The shaded area is the stability region in physical space.
The dashed line is determined by
$U=\protect\sqrt{2}\alpha/(1-\varepsilon_{\infty}/\varepsilon_{0})$ where we
took the experimental values $\varepsilon_{\infty}=4$ and $\varepsilon_0=50$.
The critical point $\alpha_c=6.8$ is represented as a full dot.}

{Fig.\,12. The same as Fig.\ 11, but now for 2D, where the critical point is
$\alpha_c=2.9$.
From (Verbist {\it et al.}, 1990).}

{Fig.\,13. States of a model two-site, two-electron system where the
electrons are coupled to the inter-atomic coordinate
from (Fisher {\it et al.}, 1989);
reproduced by courtesy of the Insitute of Physics Publishing.}

{Fig.\,14. Hall coefficient of YBa$_2$Cu$_3$O$_{7-\delta}$
for various degrees of reduction: $\delta$ = 0.05 ($\bigtriangleup$),
0.19 ($\bullet$), 0.23 ($\diamond$), 0.39 ($\circ$)
(Carrington {\it et al.}, 1993) in comparison with theory.
From (Alexandrov {\it et al.}, 1994b);
reproduced by courtesy of the Elsevier Science Publishers.}

{Fig.\,15. Resistivity of YBa$_2$Cu$_3$O$_{7-\delta}$
for various degrees of reduction in the same denotations as in Fig.\,14
(Carrington {\it et al.}, 1993) in comparison with theory.
$T^*$ stands for the temperature at which the
change in the slope of resistivity occurs.
From (Alexandrov {\it et al.}, 1994b);
reproduced by courtesy of the Elsevier Science Publishers.}

\end{document}